%% file: ppww_jhepstyle.tex
\documentclass[11pt,a4paper]{article}
\usepackage{jheppub}
\usepackage{feynarts}
\usepackage{graphics}
\usepackage{amsmath}
\usepackage{rotating}
\usepackage{url}
\include{setup_jhep}

\title{Electroweak corrections to W-boson pair
      production at the LHC}

\author[a]{Anastasiya Bierweiler,}
\author[a]{Tobias Kasprzik,}
\author[a]{Johann H.~K\"uhn,}
\author[b]{Sandro Uccirati}

\affiliation[a]{ 
Karlsruhe Institute of Technology (KIT), 
Institut f\"ur Theoretische Teilchenphysik,\\
D-76128 Karlsruhe, Germany}
\affiliation[b]{Universit\"at W\"urzburg, 
Institut f\"ur Theoretische Physik und Astrophysik,\\ 
D-97074 W\"urzburg, Germany}

\emailAdd{kasprzik@particle.uni-karlsruhe.de}
\date{\today}

\abstract{Vector-boson pair production ranks among the most important
Standard-Model benchmark processes at the LHC, not only in view of
on-going Higgs analyses. These processes may also help to gain a deeper
understanding of the electroweak interaction in general, and to test the
validity of the Standard Model at highest energies.  In this work, the
first calculation of the full one-loop electroweak corrections to
on-shell W-boson pair production at hadron colliders is presented. We
discuss the impact of the corrections on the total cross section as well
as on relevant differential distributions.  We observe that corrections
due to photon-induced channels can be amazingly large at energies
accessible at the LHC, while radiation of additional massive vector
bosons does not influence the results significantly.
\\
\vspace{1cm}
\begin{flushright}
\emph{SFB/CPP-12-58\\
TTP12-027\\
LPN12-088}
\end{flushright}}

\begin{document}
\maketitle 
\flushbottom

\section{Introduction}
%%%%%%%%%%%%%%%%%%%%%%%%% jk
With the recent start of LHC operation hard scattering processes became
accessible with parton energies up to several TeV and, indeed, jet-jet
invariant masses up to 5 TeV have been
observed~\mbox{\cite{Chatrchyan:2011ns,Aad:2011fc}} in the first round
of data taking. With the significant increase of the integrated
luminosity and the doubling of the beam energy anticipated in the next
few years, electroweak processes such as the production of lepton or
gauge-boson pairs with invariant masses of several TeV will become
accessible in the near future. These reactions may on the one hand allow
for precise tests of the Standard Model (SM) and on the other hand
potential deviations may point to ``physics beyond the SM''. Indeed, the
observation of anomalous couplings of quarks, leptons or gauge bosons
might well be the first signal of ``New Physics''. As an alternative one
may look for peaks or distortions in the W, Z or fermion spectra
resulting from the decays of new massive particles. Clearly, any
well-founded claim of ``physics beyond the SM'' must be based on precise
measurements combined with similarly precise calculations, possibly at
the level of several percent.
 
Obviously this requires theory predictions which must include
next-to-leading order (NLO), possibly even next-to-next-to-leading order
(NNLO) perturbative QCD corrections.  Considering the smallness of the
weak coupling, $\alpha_{\rm w}$, the need for electroweak (EW)
corrections is less obvious. In a first step one might try to absorb the
dominant corrections arising from the running of the fine-structure
constant from low scales to $M_{\rm Z}$ and from the
$\rho$-parameter~\cite{Ross:1975fq} in properly chosen effective
couplings. However, this approach is only justified for energies of
order $M_{\rm W}$ or $M_{\rm Z}$. As is well known from earlier
investigations \cite{Kuhn:1999de,Fadin:1999bq, Ciafaloni:2000df,
  Kuhn:1999nn, Melles:2000gw, Melles:2000ia, Denner:2003wi, Kuhn:2001hz,
  Feucht:2004rp, Jantzen:2005xi, Jantzen:2006jv, Jantzen:2005az,
  Denner:2000jv, Chiu:2009yx, Beenakker:2000kb, Beenakker:2001kf}, EW
corrections increase with the squared logarithm of the energy, and may
reach several tens of percent for energies accessible at the LHC. In
view of their strong dependence both on energy and scattering angle,
they will induce significant distortions in transverse-momentum and
rapidity distributions and consequently may well mimic ``New Physics''
and hence must be carefully taken into account.

Most of the investigations along these lines have concentrated on the
issue of Sudakov logarithms, i.e.\ terms that are at $n$ loops enhanced
proportional to $\alpha_{\rm w}^n \log^{2n-m}\big(s/M^2_{\rm W}\big)$
with \mbox{$m=0,1,\ldots,2n$}, corresponding to leading, next-to-leading
etc.\ logarithmic (N$^m$LL) enhancement. This line of research was
motivated by the necessity of including at least the dominant two-loop
terms, once one-loop corrections exceed the $20-30\%$ level.  Using
evolution equations, originally derived in the context of QED
\cite{Gribov:1966hs,Gribov:1970ik,Lipatov:1988ii,DelDuca:1990gz} and
QCD, four-fermion processes have been studied up to N$^4$LL
\cite{Kuhn:1999nn, Kuhn:2001hz,Feucht:2004rp,Jantzen:2005xi,
  Jantzen:2006jv}, W-pair production in electron--positron and
quark--antiquark annihilation up to
N$^3$LL~\cite{Kuhn:2007ca,Kuhn:2011mh}.  Employing diagrammatic methods
or the framework of soft-collinear effective field theory most of these
results were confirmed in the NLL and NNLL
approximation~\cite{Melles:2000gw,Denner:2003wi,Denner:2000jv,
  Fuhrer:2010eu}.

As stated above, the intermediate energy region, up to approximately
1~TeV, will be explored with high statistics, and the complete one-loop
corrections may become relevant. This potentially includes real photon
radiation, virtual photon exchange and terms suppressed by $M_{\rm
  W}^2/\hat{s}$, where $\hat{s}$ denotes the partonic center-of-mass
(CM) energy squared. Up to now these one-loop corrections have been
evaluated for various SM benchmark processes, including
$\mathrm{t}\bar{\mathrm{t}}$ production~\cite{Kuhn:2006vh,
  Bernreuther:2006vg, Moretti:2006nf} (for earlier studies see
ref.~\cite{Beenakker:1993yr}), $\mathrm{b}\bar{\mathrm{b}}$
production~\cite{Kuhn:2009nf}, SM-Higgs
production~\cite{Ciccolini:2003jy, Denner:2011id,
  Ciccolini:2007ec,Actis:2008ug} and decay~\cite{Passarino:2007fp,
  Kniehl:1990mq, Kniehl:1991xe, Kniehl:1991ze}, and dijet production,
where the purely weak corrections are
known~\cite{Moretti:2006ea,Dittmaier:2012kx}. For the Drell--Yan
process, the one-loop EW corrections have been studied in great detail
by many authors (see, e.g., Ref.~\cite{Kasprzik:2011ds} and references
therein) as well as for gauge-boson plus jet production, the latter in
the on-shell approximation~\cite{Kuhn:2004em,Kuhn:2005az,
  Kuhn:2005gv,Kuhn:2007qc, Kuhn:2007cv,Hollik:2007sq} and including W-
or Z-decays to leptons with all off-shell effects consistently taken
into account~\cite{Denner:2009gj,Denner:2011vu}.

So far, EW corrections to vector-boson pair production at the LHC are
only known in the high-energy
approximation~\cite{Accomando:2001fn,Accomando:2004de,
  Accomando:2005ra}, while the full one-loop EW corrections are not yet
available, despite their great phenomenological interest. These
processes contribute the most important irreducible background to the
production of a SM Higgs boson in the intermediate mass range. A
profound theoretical understanding of the underlying physics will also
allow a precise analysis of the non-abelian structure of the EW sector,
in particular of the vector boson self interactions. At the LHC these
will be explored with high precision, exploiting the high luminosity in
combination with the highest possible CM energies. Moreover,
vector-boson pair production processes are well suited for putting
experimental constraints on the existence of anomalous trilinear and
quartic gauge couplings, since their phenomenological effects are
expected to be sizable at large invariant masses of the vector-boson
pairs accessible at the LHC. Considering searches for supersymmetry
(SUSY) at the LHC, all vector-boson pair production channels constitute
important backgrounds to signals with leptons and missing transverse
energy, as e.g.\ predicted for chargino--neutralino pair production.

During recent years, a great effort has been made to push the accuracy
of the theoretical predictions to a new level. The NLO QCD corrections,
including the leptonic decays of the vector bosons, have been studied by
several authors and are implemented in Monte Carlo
programs~\cite{Ohnemus:1991gb, Frixione:1993yp, Ohnemus:1994qp,
  Ohnemus:1994ff, Dixon:1998py, Dixon:1999di, DeFlorian:2000sg,
  Campbell:1999ah, Campbell:2011bn}.  The results have been matched with
parton showers and combined with soft gluon
resummations~\cite{Frixione:2006he} to improve the predictions for
vector bosons produced at small transverse momenta. The loop-induced
channels $\mathrm{gg} \to V_1V_2$, formally a second order effect in
QCD, nevertheless play an important role at the LHC due to the
enhancement of the gluon luminosity and have also been studied
extensively~\cite{Glover:1988rg, Kao:1990tt, Duhrssen:2005bz,
  Binoth:2006mf, Campbell:2011cu}. Specifically, concerning background
estimates to Higgs production at the LHC, the contribution of the
gluon-induced channel to W-boson pair production amounts to 30\% after
experimental cuts~\cite{Binoth:2006mf}.

As stated before, EW corrections to gauge-boson pair production at
hadron colliders have only been evaluated in the high energy limit
\cite{Accomando:2001fn,Accomando:2004de, Accomando:2005ra}.  This is in
contrast to the reaction \mbox{$\mathrm{e^+e^-}\to
  \mathrm{W}^-\mathrm{W}^+$}, where the complete one-loop calculation
has long been available \cite{Lemoine:1979pm, Bohm:1987ck,
  Fleischer:1987xa, Beenakker:1993tt}, and, following the demands of LEP
experiments, the full one-loop correction for electron--positron
annihilation into four fermions (including resonant and non-resonant
amplitudes) has been calculated~\cite{Beenakker:1998gr,Jadach:2001mp,
  Denner:2000bj, Denner:2005es}.

To arrive at predictions which satisfy the needs of the next round LHC
experiments, we embark on the full one-loop corrections for W-pair
production in proton--proton collisions, where the gauge bosons are
treated as stable particles. For completeness we also recalculate the
well-known NLO QCD corrections and, furthermore, discuss two competing
processes: W-pair production through $\gamma\gamma$ collisions, and
through the quark-loop-induced gluon-fusion process. Employing the
MRST2004QED PDF set~\cite{Martin:2004dh} for the $\gamma\gamma$
luminosity, we observe a surprisingly large contribution at large
invariant W-pair mass and a pronounced peaking at small scattering
angles. In fact, for small angles and high energies this parametrically
suppressed reaction is comparable to the $q\bar{q}$ induced
reaction. Even for large angles (i.e.\ large $p_{\rT}$) it still amounts
to order of 5\% and is therefore comparable with the genuine EW
corrections. Gluon fusion, in contrast, exhibits a fairly smooth angular
distribution and is typically of order 5 to 10 percent in the whole
kinematic region of interest. Obviously all these effects must be taken
into consideration.

The layout of this paper is as follows: In Section~\ref{se:details} we start with a
qualitative discussion of W-pair production at leading order (LO). We
introduce the kinematic variables and recall the dominant features of W
production: the dependence on the W transverse momentum, W-pair
invariant mass and rapidity.  These results will illustrate the large
kinematic region accessible at the LHC and the marked differences
between the $q\bar{q}$-, $\mathrm{gg}$- and $\gamma\gamma$-induced
channels, as far as transverse-momentum and angular distributions are
concerned.  Section~\ref{se:corrections} will be devoted to a detailed discussion of
one-loop EW corrections to the reaction $q\bar{q}\to
\mathrm{W}^-\mathrm{W}^+$. We specify the renormalization scheme, our
input parameters, the treatment of QED corrections with the associated
soft and collinear singularities, their absorption in the PDFs, and the
numerical evaluation of hard photon radiation.  As stated above the rate
of \mbox{W-pair} production from $\gamma\gamma$ collisions is
surprisingly large and the resulting rates and distributions are studied
in Section~\ref{se:numres}.  The cross sections for $q\bar{q} \to
\mathrm{W}^-\mathrm{W}^+$ including EW and QCD corrections are
convoluted with PDFs, and predictions for various distributions are
shown and compared with those from $\gamma\gamma$ and gluon-fusion
processes.  For future tests and cross checks tables are presented which
precisely specify the results at various kinematic points.  In
Section~\ref{se:real}, we discuss the phenomenological impact of
additional massive vector bosons in the final state.
Section~\ref{se:concl} contains our conclusions.

\section{W-pair production: the leading order}
\label{se:details}

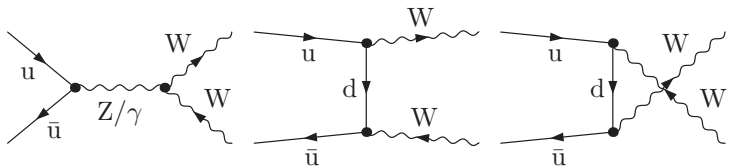
\begin{figure}[]
\begin{center}
\input{diags/diaguuWW.tex}
\end{center}
\vspace{-0.8cm}
\caption{\label{fi:tree} 
Tree-level diagrams for the LO process $\mathrm{u} \bar{\mathrm{u}} \to
\mathrm{W}^-\mathrm{W}^+$.} 
\end{figure}

At lowest order, $\mathcal{O}(\alpha^2)$, W-boson pair production at the
LHC is dominated by quark--antiquark annihilation,
\begin{equation}\label{details:LO}
 q_i\bar{q}_j \to \mathrm{W}^-\mathrm{W}^+ \,, 
\end{equation}
where the $q_k$ denote one of the light quark flavours, $q_k =
\mathrm{u,\,d,\,s,\,c}$ and $\mathrm{b}$. The corresponding diagrams are shown
in Fig.~\ref{fi:tree} for the $\Pu\Pub$ channel.
\begin{figure}[]
\begin{center}
\input{diags/diaguuWWphot.tex}
\end{center}
\vspace{-0.8cm}
\caption{\label{fi:phot} 
Tree-level diagrams for the LO processes $\gamma\gamma \to \mathrm{W}^-\mathrm{W}^+$.}
\end{figure}
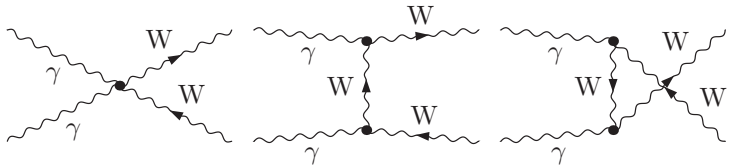
\begin{figure}[]
\begin{center}
\includegraphics[width = 1.0\textwidth]{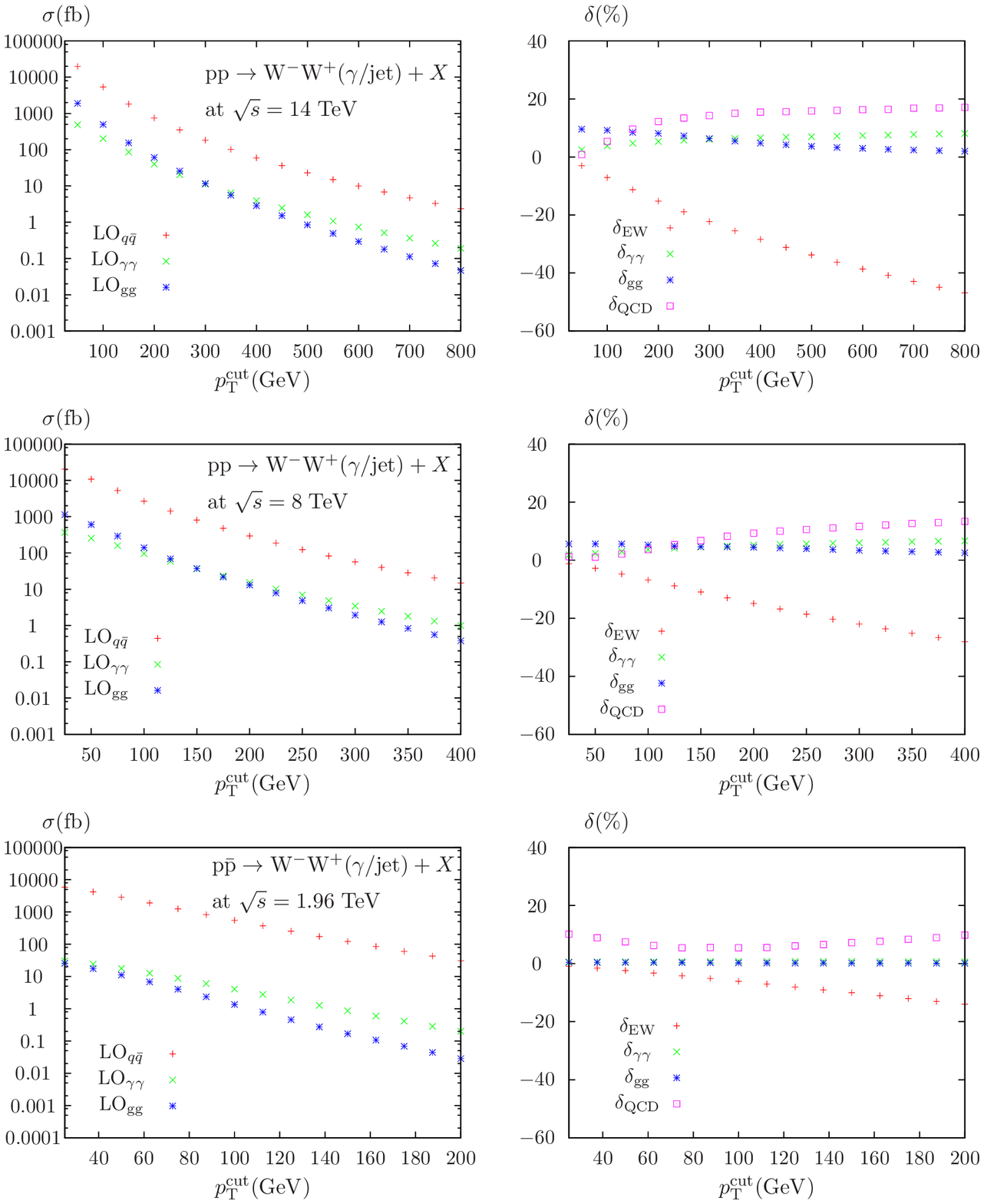}
\end{center}
\caption{\label{fi:totcs_ptw} Left: Total cross sections as a function
  of the cut on the W-boson transverse momentum at the LHC14 (top), LHC8
  (center) and the Tevatron (bottom). The corresponding relative rates
  w.r.t.\ to the $q\bar{q}$ channel are shown on the r.h.s., together
  with the rates of QCD and EW corrections. See text for details.}
\end{figure}
\begin{figure}[]
\begin{center}
\includegraphics[width = 1.0\textwidth]{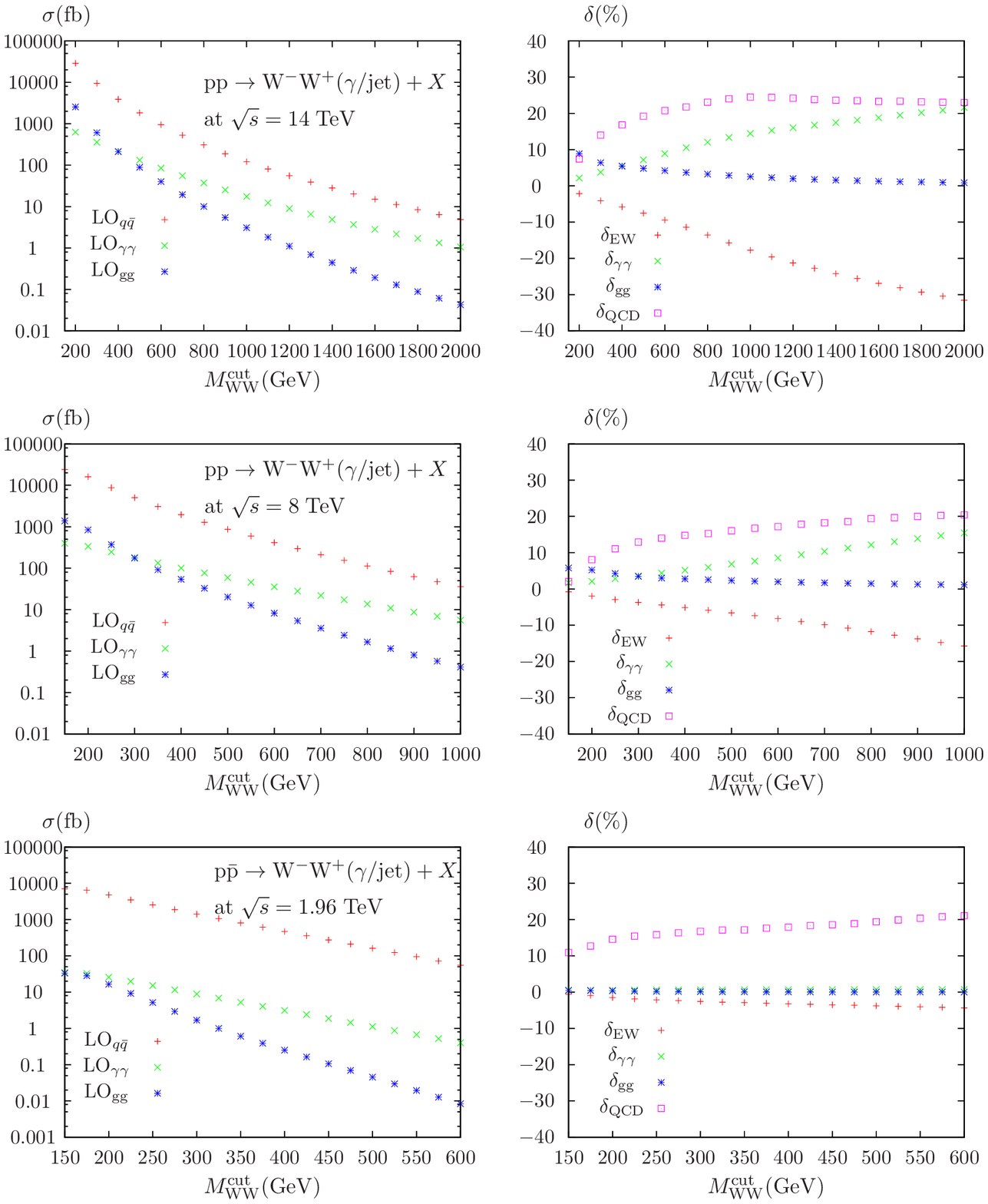}
\end{center}
\caption{\label{fi:totcs_mww} Left: Total cross sections as a function
  of the cut on the WW-invariant mass at the LHC14 (top), LHC8 (center)
  and the Tevatron (bottom). The corresponding relative rates w.r.t.\ to
  the $q\bar{q}$ channel are shown on the r.h.s., together
  with the rates of QCD and EW corrections. See text for details.}
\end{figure}

\begin{figure}[]
\begin{center}
\includegraphics[width = 1.0\textwidth]{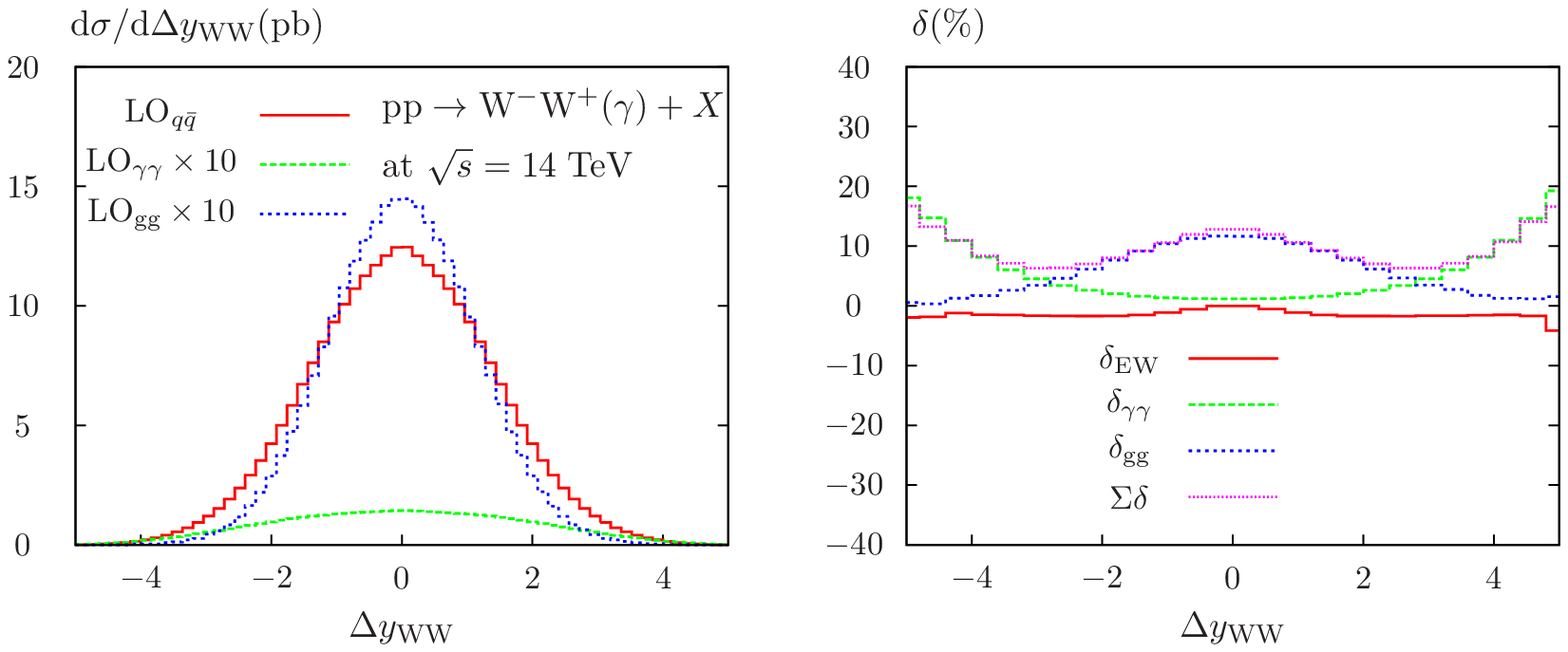}
\includegraphics[width = 1.0\textwidth]{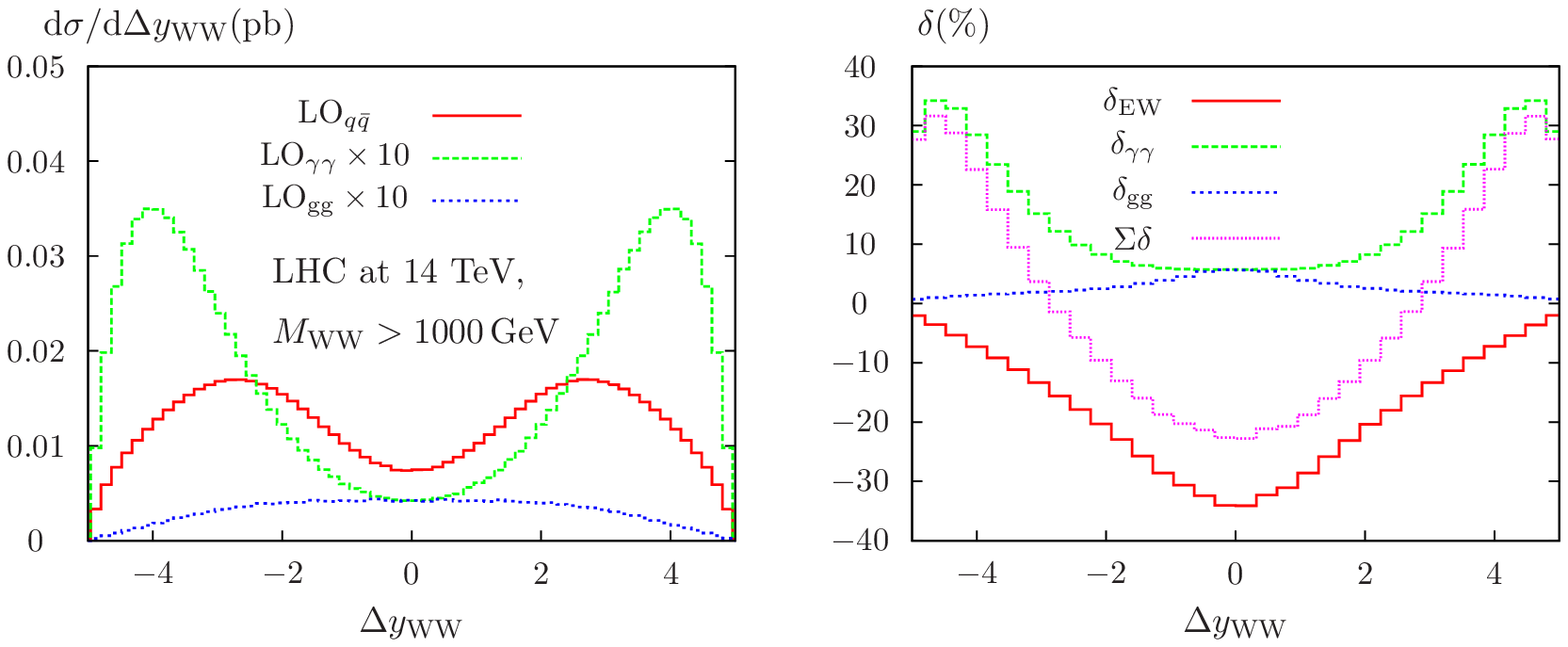}
\end{center}
\caption{\label{fi:dy_LO} Differential LO cross sections for the W-boson
  rapidity gap with default cuts (top) and with a minimal invariant mass
  of 1000~GeV (bottom) at the LHC14. On the right-hand-side, the
  corresponding relative rates due to photon- and gluon-induced
  channels w.r.t.\ the $q\bar{q}$-contributions are
  shown, as well as the EW corrections. See text for details.}
\end{figure} 

\begin{figure}[]
\begin{center}
\includegraphics[width = 1.0\textwidth]{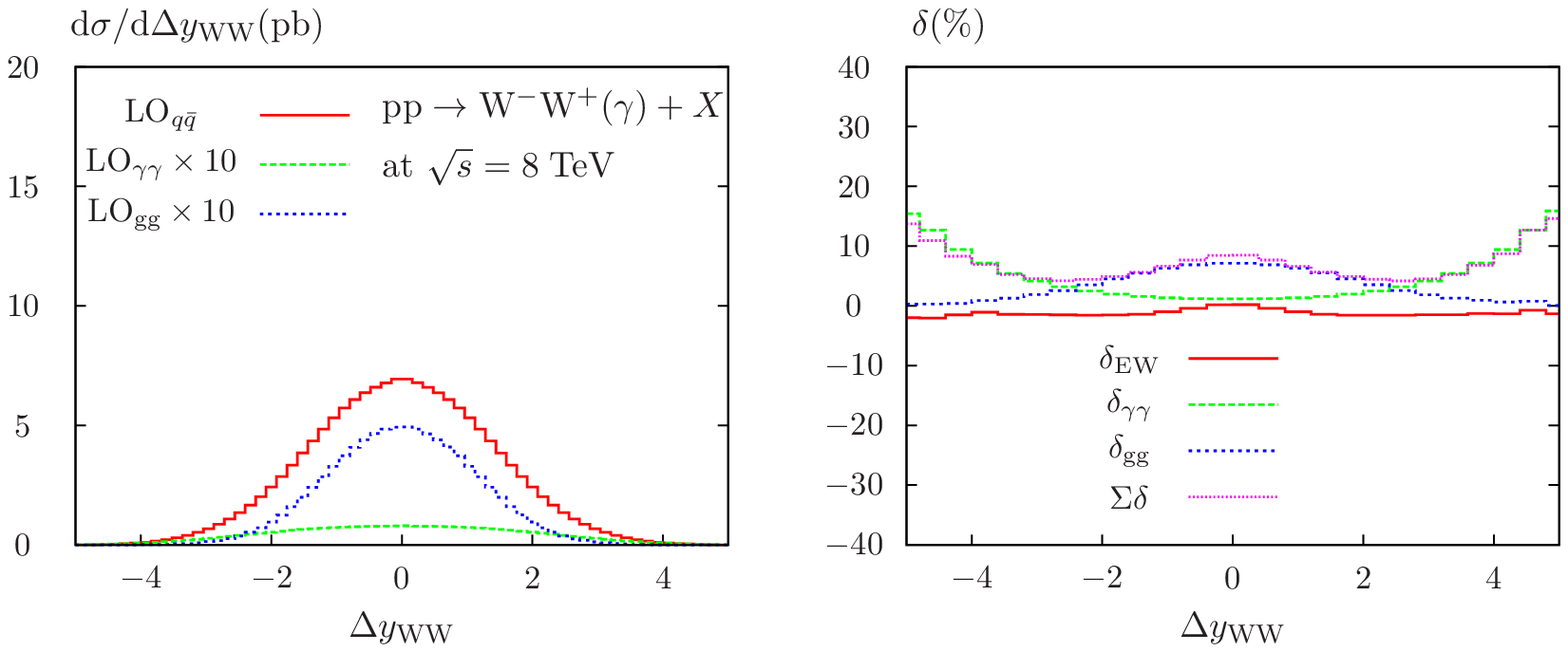}
\includegraphics[width = 1.0\textwidth]{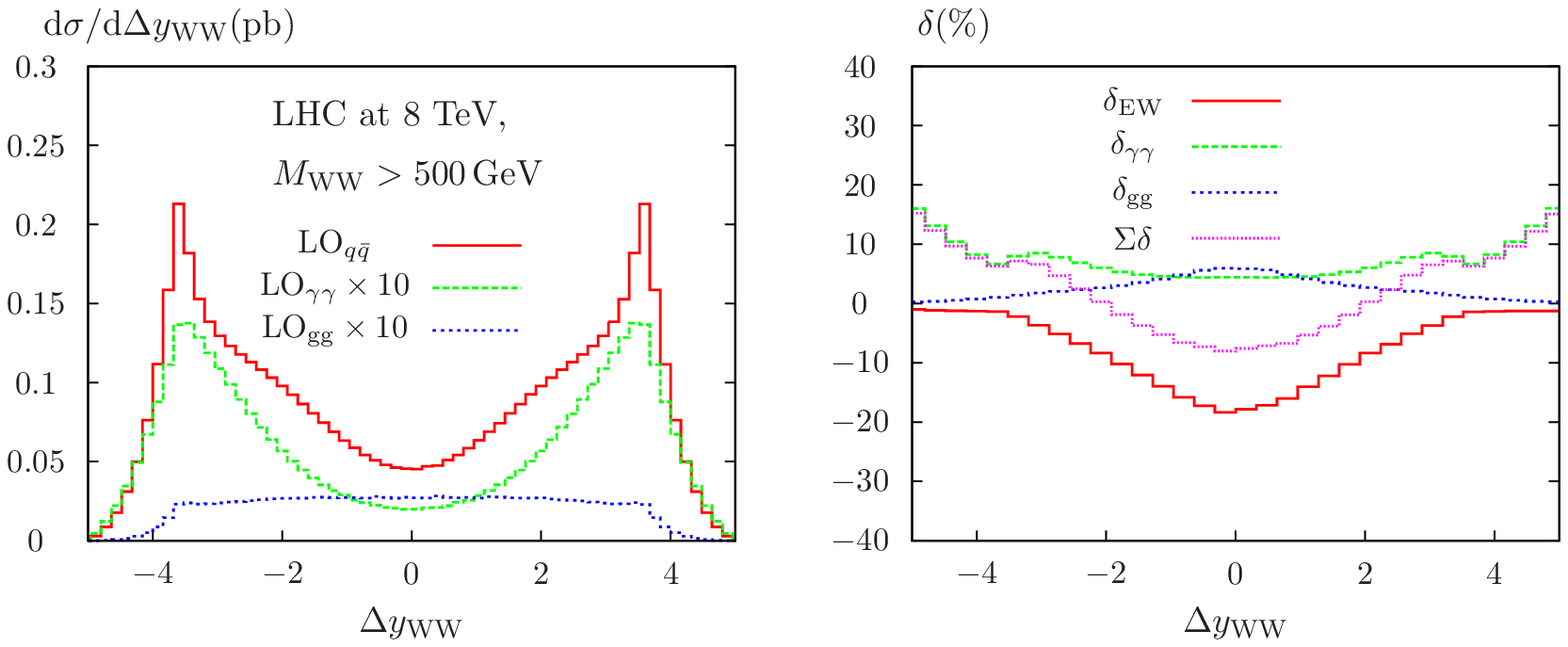}
\end{center}
\caption{\label{fi:dy_LO_L8} Differential LO cross sections for the W-boson
  rapidity gap with default cuts (top) and with a minimal invariant mass
  of 500~GeV (bottom) at the LHC8. On the right-hand-side, the
  corresponding relative rates due to photon- and gluon-induced
  channels w.r.t.\ the $q\bar{q}$-contributions are
  shown, as well as the EW corrections. See text for details.}
\end{figure} 
 Since there is a finite probability of finding a photon in a proton
we in addition take into account the photon-induced tree-level
contributions
\begin{equation}\label{details:gaga}
  \gamma\gamma \to \mathrm{W}^- \mathrm{W}^+
\end{equation}
(see Fig.~\ref{fi:phot}). In the numerical analysis the MRST2004QED PDF
set~\cite{Martin:2004dh} is used for the photon density. Although
suppressed by two photon-PDF factors, this channel may give rise to
potentially large contributions at high invariant masses due to
  peculiar features of the corresponding partonic cross section, as will
be pointed out later.

A detailed phenomenological discussion of the specific properties of
process~\eqref{details:gaga} at photon colliders can be found in
Ref.~\cite{Denner:1995jv}, where also the computation of the full
corresponding EW corrections in the SM was presented. Also the
photon-induced contributions to W-pair production in elastic
hadron--hadron scattering have been studied in an equivalent-photon
approximation~\cite{Archibald:2008zzb, Pierzchala:2008xc}, assuming a
tagging of the forward-scattered hadrons. In our approach, however, we
quantify for the first time the purely inelastic contributions which
cannot be separated from the dominating $q\bar{q}$-induced
contributions.

 For completeness, we also re-calculate the effects due to the
loop-induced channel
\begin{equation}\label{details:gg}
  \mathrm{gg} \to \mathrm{W}^-\mathrm{W}^+ \,,
\end{equation}
which adds a relative correction of about $10\%$ to the LO
$q\bar{q}$ cross section~\cite{Kao:1990tt}.

Due to the large CM energies accessible at the Tevatron and the LHC, we
can safely neglect the lepton and light quark masses (i.e.\ all but the
top-quark mass) unless they are needed to regularize infrared (IR)
singularities in the real and virtual contributions, as will be
discussed later. However, the $\Pb\Pbb$ channel receives contributions
from top-quark exchange in the $t$ channel that for consistency have to
be taken into account in the Born approximation and in the NLO calculation. In
this work we assume a block-diagonal CKM matrix with $V_{\rm tb}
=1$. Therefore, the cross sections~\eqref{details:LO} are
proportional to $\sum_{k=1}^2 V_{ik}V^*_{kj} = \delta_{ij} \;
\mbox{for}\; i,j=1,2$, and it is sufficient to only consider
flavour-diagonal processes. Consequently, at the parton level the
complete one-loop calculation has to be carried out independently for
the following three different channels,
\begin{subequations}
  \begin{eqnarray}
    \mathrm{u} \,\bar{\mathrm{u}} &\to& \mathrm{W}^-\mathrm{W}^+\,,  \\
    \mathrm{d} \,\bar{\mathrm{d}} &\to& \mathrm{W}^-\mathrm{W}^+\,,  \\
    \mathrm{b} \,\bar{\mathrm{b}} &\to& \mathrm{W}^-\mathrm{W}^+\,.  
  \end{eqnarray}
\end{subequations}

Let us, in a first step, evaluate the LO prediction for $q\bar{q}$-,
$\gamma\gamma$- and $\mathrm{gg}$-induced processes. The cross sections for
W-pairs as a function of $p_{\rT}^{\mathrm{cut}}$, corresponding to a
cut on the transverse momenta of the W bosons, and as a function of
$ M_{\PW\PW}^{\mathrm{cut}}$, corresponding to a cut on the mass of the
W-pair, are shown in Figs.~\ref{fi:totcs_ptw} and~\ref{fi:totcs_mww}, 
respectively. The input values and setup we use for our computation will
be specified in Section~\ref{se:SMinput}, together with the rapidity and
$p_{\rT}$ cuts employed throughout. Given an integrated luminosity
of $20~\mathrm{fb}^{-1}$ at 8~TeV and $200~\mathrm{fb}^{-1}$ at
$14~\TeV$, transverse momenta (invariant masses) up to $0.35~\TeV$
($1~\TeV$) can be explored at 8 TeV and up to $0.75~\TeV$ ($2~\TeV$)
at 14~TeV. Here, we assume that events with one W boson decaying into
$\mathrm{e\nu_e}$ or $\mu\nu_{\mu}$ and one decaying hadronically can be detected
with 30\% efficiency and require more than 100 detected events.

When comparing the $p_{\rT}^{\mathrm{cut}}$ and
$M_{\PW\PW}^{\mathrm{cut}}$ dependence of $q\bar{q}$-, $\gamma\gamma$-
and $\mathrm{gg}$-induced processes, marked differences can be
observed. For large $M_{\PW\PW}^{\cut}$ the rates of $\gamma\gamma$- and
$q\bar{q}$-induced reactions are quite comparable. However, the cut on
$p_{\rT}$ reduces the relative effect of the photon-induced channels
significantly. This is shown on the right hand sides of the respective
figures, where the cross sections of the $\mathrm{gg}$- and
$\gamma\gamma$-induced reactions relative to the $q\bar{q}$-induced
process (denoted as $\delta_{\mathrm{gg}}$ and $\delta_{\gamma\gamma}$,
respectively) are plotted. Let us first concentrate on the LHC at 14 TeV
(LHC14). The relative rate for W-pairs from gluon fusion amounts to less
than 10\% for small $p_{\rT}$ and decreases with increasing $p_{\rT}$
(Fig.~\ref{fi:totcs_ptw}). The parametric suppression of $\mathrm{gg}\to
\mathrm{W^-W^+}$ by $\alpha_s^2$ is compensated by the large gluon
luminosity for small $\tau = \hat{s}/s$ which, however, dies out with
increasing $\tau$. A qualitatively similar behaviour is observed for a
cut on $M_{\PW\PW}$ (Fig.~\ref{fi:totcs_mww}). The relative importance
of the photon-induced process, in contrast, increases with increasing
$p_{\rT}^{\cut}$, and even more so with $M_{\PW\PW}^{\cut}$. This
behaviour can be traced to the energy dependence of the cross section
for the partonic subprocess $\gamma\gamma\to \mathrm{W^-W^+}$, which
approaches a constant value in the high-energy limit,
$\hat{\sigma}_{\gamma\gamma}\approx 8\pi\alpha^2/M_{\mathrm{W}}^2$, with
an extremely strong peaking in the forward and backward directions. The
cut on $p_{\rT}$, when compared to the one on $M_{\PW\PW}$ thus leads to
a significantly stronger reduction of the event rate. Nevertheless, a
10\% admixture of photon-induced W-pairs remains for $p_{\rT}$ above
800~GeV. Qualitatively similar statements are applicable to the LHC at
8~TeV (LHC8). However, the relative contributions of both
\mbox{$\mathrm{gg}$- and $\gamma\gamma$-induced} reactions are
significantly smaller, and become completely negligible for the
Tevatron. In the right three plots of Figs.~\ref{fi:totcs_ptw}
and~\ref{fi:totcs_mww} we also anticipate some of the results on
one-loop QCD ($\delta_{\QCD}$) and EW ($\delta_{\EW}$) corrections. For
transverse momenta around 800~GeV, accessible at LHC14, large negative
EW corrections amount to nearly $-50\%$ and are the main subject of this
paper. In fact, they are quite comparable to the positive QCD
corrections, which are displayed in the same figures. Note that the
latter strongly depend on the cuts on real jet radiation, as will be
discussed at the end of Section~\ref{se:QCD}.

As indicated above, the difference between the $p_{\rT}^{\cut}$ and
$M_{\PW\PW}^{\cut}$ dependence of the cross sections can be traced to
the different angular distributions of the W bosons in the W-pair rest
frame, which for $\gamma\gamma$ is strongly peaked in the forward and
backward directions. The marked differences in the angular distributions
are illustrated in Fig.~\ref{fi:dy_LO}, where the distributions in the
rapidity difference $\Delta y_{\PW\PW} =
y_{\mathrm{W}^-}-y_{\mathrm{W}^+}$ (which, for fixed $M_{\PW\PW}$,
corresponds to the angular distribution in the W-pair rest frame) are
shown for the mass intervals $[2M_{\PW},\infty]$ and
$[1000\;\GeV,\infty]$, respectively. In the first case, the cross
section is dominated by central events with W-pairs of low invariant
mass and relatively isotropic angular distributions. All three
components are peaked at small $\Delta y_{\PW\PW}$. Considering their
ratios, displayed on the r.h.s.\ one observes the stronger preference of
the $\mathrm{gg}$ process for small $\Delta y_{\PW\PW}$ and, conversely,
the enhancement of the $\gamma\gamma$ process for large $\Delta
y_{\PW\PW}$. Considering events with large invariant mass, $M_{\PW\PW} >
1~\TeV$, the distribution of the three components exhibits a drastically
different behaviour. The rate of the gluon-fusion process is fairly
small and its angular distribution does not exhibit any pronounced
structure, resulting in a flat rapidity distribution. As expected,
W-pair production is dominated by $q\bar{q}$ annihilation with its
strong peaking at small $\hat{t}$, i.e.\ in the forward and backward
directions.  This is reflected in the peaks of $\Delta y_{\PW\PW}$ around
$\pm 2.5$. The contribution from $\gamma\gamma$ fusion remains sizable,
with a very pronounced peaking for large rapidity difference,
corresponding to a highly anisotropic angular distribution in the
$\mathrm{WW}$ rest frame. The EW corrections exhibit a completely
different behaviour. For large $\Delta y_{\PW\PW}$, corresponding to
small $\hat{t}$ they are negative but less than 10\%. In the Sudakov
regime, however, corresponding to small $\Delta y_{\PW\PW}$ and thus
large $\hat{t}$, the Sudakov effect comes into play and negative
corrections of more than 30\% are observed. In total, the sum of these
corrections, denoted $\Sigma\delta$ in the right displays of
Fig.~\ref{fi:dy_LO}, will lead to a dramatic distortion of $\rd \sigma /
\rd \Delta y_{\PW\PW}$ with corrections varying between $+30\%$ and
$-30\%$ for large and small $|\Delta y_{\PW\PW}|$, respectively. Note
that such a behaviour could well be misinterpreted as a signal for
anomalous couplings.\footnote{The interplay of logarithmic EW
  corrections and anomalous trilinear gauge-boson couplings has been
  studied in Ref.~\cite{Accomando:2005xp} for WZ and WW production.} A
less pronounced, but qualitatively similar behaviour is observed for
LHC8 with the mass intervals $[2 M_{\mathrm{W}},\infty]$ and $[500\;{\rm
  GeV},\infty]$ (Fig.~\ref{fi:dy_LO_L8}).

\section{Radiative corrections}
\label{se:corrections}
In addition to the tree-level contributions, a full
$\mathcal{O}(\alpha^3)$ analysis of the W-boson pair production cross
section requires the inclusion of the one-loop EW corrections to the
$q\bar{q}$-induced processes as well as
the contributions due to radiation of one additional bremsstrahlung
photon.  Thus, the corresponding total partonic cross section at NLO may
be written as 
\begin{equation}
  \hat{\sigma}^{ij\to \WW(\gamma)}_{\NLO} = \hat{\sigma}^{q\bar{q}\to
    \WW}_{\LO} +\hat{\sigma}^{\gamma\gamma\to  \WW}_{\LO}
  +\hat{\sigma}^{\mathrm{gg} \to  \WW}_{\LO} + 
  \hat{\sigma}^{q\bar{q}\to \WW}_{\mathrm{virt}} + \hat{\sigma}^{q\bar{q}\to  \WW\gamma}_{\mathrm{\LO}}\,,
\end{equation}
with $q = \mathrm{u,\,d,\,s,\,c,\,b}$, and the different partonic
contributions are given by
\begin{eqnarray}
  \hat{\sigma}^{\gamma\gamma\to  \WW}_{\LO} &=& \frac{1}{N_{\gamma\gamma}}\frac{1}{2\hat{s}}\int\rd\Phi(\PW^-\PW^+) \sum_{\mathrm{pol}}
  |\M^{\gamma\gamma\to \WW}_0|^2\,, \\
 \hat{\sigma}^{\mathrm{gg} \to  \WW}_{\LO} &=&
 \frac{1}{N_{\mathrm{gg}}}\frac{1}{2\hat{s}}\int\rd\Phi(\PW^-\PW^+)  \sum_{\mathrm{col}} \sum_{\mathrm{pol}}
  |\M^{\mathrm{gg} \to \WW}_1|^2\,, \\   
  \hat{\sigma}^{q\bar{q}\to  \WW}_{\LO} + \hat{\sigma}^{q\bar{q}\to \WW}_{\mathrm{virt}} &=&
  \frac{1}{N_{q\bar{q}}}\frac{1}{2\hat{s}}\int\rd\Phi(\PW^-\PW^+)
  \nonumber \\
 &&\hspace{-2cm}\times \sum_{\mathrm{col}}
  \sum_{\mathrm{spin}} \sum_{\mathrm{pol}}
  \left[|\M^{q\bar{q} \to \PW\PW}_0|^2 + 2 \Re \left\{(\M^{q\bar{q} \to \PW\PW}_0)^*\;\M^{q\bar{q} \to \PW\PW}_1\right\}\right]\,,\\
  \hat{\sigma}^{q\bar{q}\to  \WW\gamma}_{\mathrm{\LO}} &=&\frac{1}{N_{q\bar{q}}}\frac{1}{2\hat{s}}
  \int\rd\Phi(\PW^-\PW^+\gamma) \sum_{\mathrm{col}} \sum_{\mathrm{spin}}
  \sum_{\mathrm{pol}} |\M^{q\bar{q} \to \PW\PW \gamma}_0|^2\,.
\end{eqnarray}
Here, $\M_0$ and $\M_1$ denote the corresponding Feynman amplitudes of
the tree-level and one-loop contributions, respectively,
$\int\rd\Phi(\mathrm{final\;state})$ is the Lorentz invariant
phase-space measure of the final-state particles and $\hat{s}$ is the
partonic CM energy. The normalization factors are given by $N_{q\bar{q}}
= 36$, $N_{\gamma\gamma} = 4$ and $N_{\mathrm{gg}} = 256$.

Although the photon-induced process~\eqref{details:gaga} is considered
at leading order, we emphasize that it formally contributes at
next-to-next-to-leading order in the QED coupling constant according to
Eq.\ (5) of Ref.~\cite{Martin:2004dh}. Thus, we do not include the EW
corrections to the $\gamma\gamma$ process (including the $\gamma\gamma
\to \PW^-\PW^+\gamma$ channel) in our analysis. Since the corresponding
relative EW corrections are at the level of 10\% at high energies and
small scattering angles (see Fig.\ 9 of Ref.~\cite{Denner:1995jv}), the
related uncertainties w.r.t.\ the $q\bar{q}$ channel are never exceeding
3\% even at high invariant masses and may safely be neglected.  For a
similar reasoning we also do not take into account the real-radiation
processes $q(\bar{q})\gamma \to q(\bar{q}) \PW^-\PW^+$. The dominating
contribution from this process class, attributed to the collinear
$q(\bar{q}) \to \gamma^* q(\bar{q})$ splitting, is already included in
the definition of the photon PDFs, while the configuration corresponding
to collinear $\gamma \to q^*\bar{q}$ and $\gamma \to q\bar{q}^*$
splittings can be interpreted as a QED correction to the (anti)quark
PDF, which is estimated to be small~\cite{Roth:2004ti}. The residual
contributions, namely photon-induced WW + jet production again gives a
genuine NNLO contribution.  Moreover, including all these contributions
(which of course would be possible), to properly account for all
one-loop QED effects would require the rigorous application of the
out-dated MRST2004QED PDF set which is discouraged by the authors of
Ref.~\cite{Martin:2004dh}.

 Defining the momentum fractions $x_a$ and $x_b$ of the initial-state
 hadrons carried by the incoming partons, the hadronic NLO cross
 sections at the LHC is given by a convolution of the partonic cross
 section with the parton distribution functions (PDFs),
\begin{equation}
  \sigma_{\NLO}^{\mathrm{pp \to WW(\gamma)}} = \int_{\tau_0}^1 \rd \tau \int_\tau^1 \frac{\rd x_b}{x_b}
  \, \sum_{i,j} 
  f_{i/\mathrm{p}}(x_a,\muF^2)f_{j/\mathrm{p}}(x_b,\muF^2)\,\hat{\sigma}_{\NLO}^{ij \to
    \mathrm{WW(\gamma)}}(\tau s,\muF^2)\,,
\end{equation}
where the hadronic CM energy $s$ is related to $\hat{s}$ via $\hat{s} =
\tau s$, with $\tau = x_ax_b$.  The kinematic production threshold of a
W-boson pair in the final state is reflected in the choice of the lower
integration boundary $\tau_0 = 4 M_{\rm W}^2/s$, corresponding to a minimal
partonic CM energy of $\hat{s}_0 = \tau_0 s$, and $\muF$ denotes the
factorization scale.

We have performed two completely independent calculations of the EW
corrections, and the results displayed in
Tables~\ref{ta:totcs_def}--\ref{ta:mww_os} have been reproduced by both
of them.

\subsection{Virtual corrections and renormalization}
\label{se:virtual} 
The virtual one-loop corrections receive contributions from
self-energies, triangles and box diagrams.  In the first approach the
diagrams are automatically generated with {\tt
  Feyn\-Arts\;3.5}~\cite{Kublbeck:1990xc,Hahn:2000kx} and {\tt
  FormCalc\;6.1}~\cite{Hahn:1998yk,Hahn:2001rv} is used to calculate and
algebraically simplify the corresponding amplitudes. Afterwards, the
multiplication with the Born-level amplitude, as well as the summation
(averaging) over polarisations, spins and colours is performed
completely analytically within the {\tt FormCalc} framework. 

In an alternative approach we used the program {\tt
  QGraf}~\cite{Nogueira:1991ex} to generate Feynman diagrams. The
implementation of Feynman rules, the reconstruction of the Dirac
structure (as well as the evaluation of Dirac traces) and the
calculation of squared matrix elements is carried out analytically using
the computer algebra program {\tt FORM}~\cite{Vermaseren:2000nd}. To
avoid numerical instabilities, potentially small Gram-determinants,
which occur in the tensor reduction of particular four-point-functions
are cancelled at the analytical level using {\tt FORM}.

We use two different analytical implementations of the
Passarino--Veltman algorithm~\cite{Passarino:1978jh} (based on
\emph{Mathematica} and {\tt FORM}, respectively) to reduce tensor
coefficients to scalar integrals. The reduction was also tested against
the numerical approach implemented in the {\tt
  Loop\-Tools\;2.5}~\cite{Hahn:1998yk,vanOldenborgh:1989wn} library,
which is used for the evaluation of the scalar one-loop integrals.

For the calculation of the gluon-induced process~\eqref{details:gg} we
use the fully automated setup of {\tt FeynArts} and {\tt FormCalc},
where the computation of the squared one-loop amplitude as well as the
summation over polarisations are carried out numerically.

The ultraviolet (UV) divergences that arise in the computation of the
one-loop diagrams are treated in dimensional regularization going from 4
to $D = 4-2 \epsilon$ space-time dimensions, where the UV divergences
appear as single poles in the small complex parameter $\epsilon$. After
adding the counterterms in a proper renormalization procedure, the poles
vanish, and the limit $\epsilon \to 0$ can be taken to obtain physical
results. 

Our results are based on the on-shell renormalization scheme
defined in the following.

\subsubsection{On-shell scheme}
Our choice for the renormalization prescription is the
on-shell (OS) renormalization scheme as specified
in Ref.~\cite{Denner:1991kt}. Instead of defining the electromagnetic
coupling constant $\alpha$ in the Thomson-limit, however, we work in the
$\GF$ scheme where $\alpha$ is derived from the Fermi-constant $\GF$ via
\begin{equation}
  \alpha_{\GF} = \frac{\sqrt{2}\GF\MW^2}{\pi}\left(1-\frac{\MW^2}{\MZ^2}\right)\,.
\end{equation}
In this scheme, the weak corrections to muon decay $\De r$ are
included in the charge renormalization constant $\delta Z_e$ by the replacement 
\begin{equation}
  \delta Z_e\big|_{\alpha(0)} \to \delta Z_e\big|_{\alpha_{\GF}} =
  \delta Z_e\big|_{\alpha(0)} - \frac{\Delta r}{2} 
\end{equation}
in the calculation of the counterterm contributions (see, e.g.,
Ref.~\cite{Dittmaier:2001ay}).  As a consequence, the EW corrections are
independent of logarithms of the light-quark masses. Moreover, this
definition effectively resums the contributions associated with the
running of $\al$ from zero to the weak scale and absorbs some leading
universal corrections $\propto\GF\Mt^2$ from the $\rho$~parameter into
the LO amplitude.

\subsection{Real corrections}
\label{se:real} 
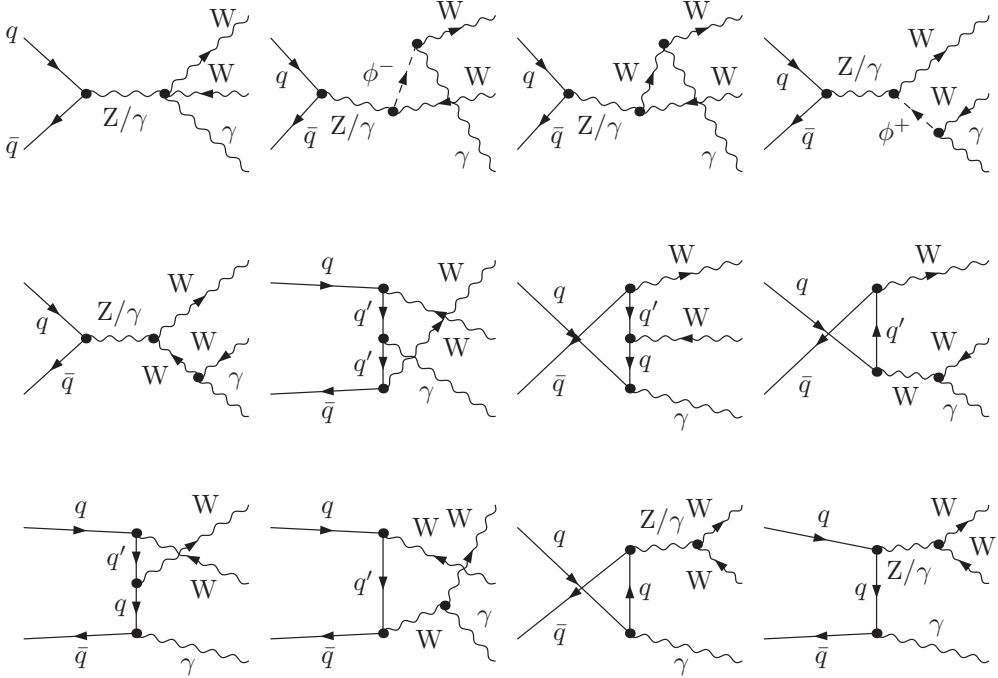
\begin{figure}
\begin{center}
\input{diags/diaguuWWreal.tex}
\end{center}
\caption{\label{fi:real} 
Generic bremsstrahlung diagrams for the process
$q\bar{q} \to \mathrm{W}^-\mathrm{W}^+\gamma$.} 
\end{figure}
In a second step, the diagrams contributing to the bremsstrahlung
amplitude $\M_0^{q\bar{q} \to \mathrm{WW}\gamma}$ (see
Fig.~\ref{fi:real}) must be considered.  In the first approach the
amplitudes are generated with {\tt FeynArts} and analytically squared
within \emph{Mathematica} using {\tt FeynCalc}~\cite{Mertig:1990an}. We
use {\tt MadGraph}~\cite{Alwall:2007st} and {\tt FormCalc} for internal
checks, however, the computational performance of the {\tt
  FeynArts/FeynCalc}-based code turns out to be more efficient.
Alternatively, the corresponding Feynman-diagrams are generated with
{\tt QGraf} and the corresponding amplitudes, as well as the squared
matrix elements, are evaluated analytically with {\tt FORM}.  In both
approaches, the numerical evaluation of the bremsstrahlung contributions
is carried out in {\tt FORTRAN} using the {\tt
  VEGAS}~\cite{Lepage:1977sw} algorithm.

In the computation of the real corrections, care has to be taken since
the phase-space integral over $|\M_0^{q\bar{q} \to
  \mathrm{WW}\gamma}|^2$ exhibits IR singularities in phase-space
regions where the photon is radiated collinear to an initial-state quark
or becomes soft, i.e.\ the energy of the photon goes to zero. We apply
the well-known technique of the two--cut-off phase-space
slicing~\cite{Harris:2001sx} to analytically carry out the phase-space
integration over the soft and collinear singularities by using small
mass regulators $m_q$ and $\lambda$ for the light quarks and the photon,
respectively, and exploiting universal factorization properties for the
squared bremsstrahlung amplitudes in the soft and collinear limit (see,
e.g., Ref.~\cite{Dittmaier:1999mb}). Accordingly, the IR singularities
appear as $\ln m_q$ and $\ln \lambda$ terms in this particular mass
regularisation scheme. Note that the hierarchy $\lambda \ll m_q$ has to
be respected carefully in the evaluation of the IR-singular terms. To
separate the soft and collinear phase-space configurations from the hard
bremsstrahlung, one imposes a cut on the photon energy, $E_\gamma <
\Delta E \ll \sqrt{\hat{s}}$ and a cut on the angle between incoming
(anti-)quark and photon, $\theta_{q\gamma} < \Delta \theta \ll 1$. The
residual phase-space integration corresponding to hard, non-collinear
photon emission, i.e.\ $E_\gamma > \Delta E$ and $\theta_{q \gamma} >
\Delta\theta$, can safely be carried out without IR regulators, allowing
for an efficient numerical evaluation using {\tt Vegas}.  Adding the
soft, collinear and hard contributions, the dependence on the slicing
parameters $\Delta E$ and $\Delta \theta$ cancels out in the computation
of IR-safe observables.\footnote{Technically, the dimensionless quantity
  $\delta_s = 2 \Delta E/\sqrt{\hat{s}}$ is used as slicing parameter
  instead of $\Delta E$.} For a more detailed description of the
phase-space slicing method as applied in our computation, including all
relevant formulae and further useful references, see
e.g.~Ref.~\cite{Ciccolini:2003jy}.

According to the Bloch--Nordsieck theorem~\cite{Bloch:1937pw}, the soft
singularities emerging from real radiation cancel against corresponding
contributions from the virtual corrections related to photon exchange
between on-shell legs in loop diagrams. However, the initial-state
collinear singularities survive and have to be absorbed in the
renormalized PDFs, where we apply the $\overline{\mathrm{MS}}$
factorization scheme as described in Ref.~\cite{Diener:2003ss}.  Since
we do not include the leptonic decays of the final-state W bosons in
this work, we do not have to deal with subtleties concerning infrared
safety arising from the event-selection procedure for photon radiation
off charged leptons in the final state.

\subsection{NLO QCD contributions}
\label{se:QCD}
In addition to the EW corrections, we have also recalculated the NLO QCD
corrections to W-boson pair production, using the same setup as detailed
above. The corresponding real corrections exhibit additional partonic
channels with one gluon in the initial state, namely
\begin{subequations}
\begin{eqnarray}
q \,\mathrm{g} &\to& \mathrm{W}^-\mathrm{W}^+\;q \,, \\
\mathrm{g} \,\bar{q} &\to& \mathrm{W}^-\mathrm{W}^+\;\bar{q}\,,
\end{eqnarray}
\end{subequations}
where the contributions $\mathrm{g}\,\mathrm{b} \to \mathrm{W}^-
\mathrm{t}^* \to \mathrm{W}^-\mathrm{W}^+ \,\mathrm{b} $ and 
$\mathrm{g}\,\bar{\mathrm{b}} \to \mathrm{W}^+ \bar{\mathrm{t}}^* \to
\mathrm{W}^+\mathrm{W}^-\,\bar{\mathrm{b}} $ formally contribute
to associated production of a W-boson and a potentially resonant
top-quark at leading order and thus have to be treated separately. We
exclude those channels from our analysis by discarding events with a
$\mathrm{b}$-quark in the final state, assuming a 100\% tagging
efficiency.

In the numerical analysis one finds that the $p_\rT$ distribution of the
W-boson is plagued by huge NLO QCD $K$-factors of a few hundreds of percent,
resulting in potentially large uncertainties in the theory predictions. To
circumvent this problem, we follow the strategy proposed in
Ref.~\cite{Denner:2009gj} and veto events with a hard jet recoiling
against a hard $\mathrm{W}$, because these signatures belong to
$\mathrm{W}+$jet production at LO rather than being a correction to
W-pair production. Technically, we discard events where the
transverse momentum of a visible jet (with
$p_{\rT,\mathrm{jet}}>15\;\mathrm{GeV}$ and $|y_{\mathrm{jet}}| < 2.5$)
is larger than half of the highest W-boson $p_{\rT}$. (The important
issue of giant QCD $K$-factors has also been discussed in
Ref.~\cite{Rubin:2010xp}.) 

\section{Numerical results}
\label{se:numres}
In this section we present numerical results for total cross sections
and differential distributions for the process $\mathrm{p\bar{p}} \to
\mathrm{W}^-\mathrm{W}^++X$ at the Fermilab Tevatron for a total CM
energy of $\sqrt{s} = 1.96\,\TeV$, as well as for the process
$\mathrm{pp} \to \mathrm{W}^-\mathrm{W}^++X$ at the CERN LHC with CM
energies of $\sqrt{s} = 8 \, \TeV$ (LHC8) and $\sqrt{s} = 14 \, \TeV$
(LHC14), respectively. We discuss in detail the phenomenological
implications of the EW corrections.  In the following the relative
corrections $\delta$ are defined through $\sigma_{\NLO} =
(1+\delta)\times\sigma_{\LO}$. The photon- and gluon-induced
channels~\eqref{details:gaga} and \eqref{details:gg}, although formally
contributing at leading order, are also considered as relative
corrections $\delta_{\gamma\gamma}$ and $\delta_{\mathrm{gg}}$,
respectively.

\subsection{Input parameters and event selection}
\label{se:SMinput}
We use the following SM input parameters for the numerical analysis, 
\begin{equation}\arraycolsep 2pt
\begin{array}[b]{lcllcllcl}
\GF & = & 1.16637 \times 10^{-5} \;\GeV^{-2}, \quad & & & & &  \\
M_{\mathrm{W}} & = & 80.398\;\GeV, & M_{\mathrm{Z}} &
= & 91.1876\;\GeV, \quad & & & \\ 
 M_\PH & = & 125\;\GeV, & m_\Pt & = & 173.4\;\GeV\,. & & & 
\end{array}
\label{eq:SMpar}
\end{equation}
In the on-shell scheme
applied in our computation, the weak mixing angle
$\cos^2\theta_{\mathrm{w}} = M_\mathrm{W}^2/M_{\mathrm{Z}}^2$ is a
derived quantity.  For the computation of the process \mbox{$q\bar{q}
  \to \mathrm{W}^-\mathrm{W}^+$} and its EW radiative corrections, we
use the MSTW\-2008\-LO PDF set~\cite{Martin:2009iq} in the LHAPDF
setup~\cite{Whalley:2005nh}. In order to consistently include
$\mathcal{O}(\alpha)$ corrections, in particular real radiation with the
resulting collinear singularities, PDFs should take in principle these
QED effects into account. Such a PDF analysis has been performed in
Ref.~\cite{Martin:2004dh}, and the $\mathcal{O}(\alpha)$ effects are
known to be small, as far as their effect on the quark distribution is
concerned~\cite{Roth:2004ti}. In addition, the currently available PDFs
incorporating $\mathcal{O}(\alpha)$ corrections~\cite{Martin:2004dh}
include QCD effects at NLO, whereas our EW analysis is LO with respect
to perturbative QCD only. For these reasons, the MSTW2008LO set is used
as our default choice for the $q\bar{q}$ process.

W-pair production through gluon fusion is considered at leading order,
hence the same leading-order set~\cite{Martin:2009iq} is
implemented. For W-pair production through the partonic subprocess
\mbox{$\gamma\gamma \to \mathrm{W}^-\mathrm{W}^+$} we use the MRST\-2004\-QED
set~\cite{Martin:2004dh}. Note that this set has been also used in
Ref.~\cite{Dittmaier:2009cr} to predict the contamination of the
Drell--Yan process by the $\gamma\gamma \to \mu^+\mu^-$ contribution.

For the computation of the QCD corrections, however, the MSTW2008NLO set
is adopted, corresponding to a value of
$\alphas(M_{\mathrm{Z}}) = 0.1202$. As 
explained above, the CKM matrix can be assumed to be diagonal. The
renormalization and factorization scales are always identified, our
default scale choice being the phase-space dependent average of the
W-boson transverse masses
\begin{equation}
\muR = \muF = \overline{m_{\rT}} =
\frac{1}{2}\left(\sqrt{M_{\PW}^2+p_{\rT,\PW^-}^2}+\sqrt{M_{\PW}^2+p_{\rT,\PW^+}^2}\right)
\end{equation}
%or the invariant mass of the
%W-boson pair, $M_{\PW\PW}= \sqrt{(p_{\PW^-}+p_{\PW^+})^2}$, 
for the evaluation of the QCD as well as the EW
corrections.\footnote{Although the relative EW corrections hardly depend
  on the choice of the factorization scale, the QCD corrections can be
  significantly stabilized using a scale which is adjusted to the
  kinematics of the underlying hard process. (For a detailed discussion
  of this important issue, see, e.g.,\ Ref.~\cite{Denner:2011vu}.)}  
A similar scale choice was taken in Ref.~\cite{Accomando:2004de} for the
computation of the EW corrections to four-lepton production at the LHC. 
In our default setup, we require a minimum transverse momentum and a
maximum rapidity for the final-state W bosons,
\begin{equation}\label{eq:defcuts}
 p_{\rT,\PW^{\pm}} > 15\;\GeV\,,\quad |y_{\PW^{\pm}}|<2.5\,,
\end{equation}
to exclude events where the bosons are emitted collinear to the initial-state
partons. As far as QCD corrections are concerned, the jet veto
introduced before is only applied to visible jets with
 \begin{equation}\label{eq:defcuts}
 p_{\rT,\jet} > 15\;\GeV\,,\quad |y_{\jet}|<2.5
\end{equation}
to ensure infrared safety.

\subsection{Integrated cross sections}
\label{se:CSresults}
 We start the discussion of our numerical
  results with the presentation of integrated cross sections at the LHC
  and Tevatron, evaluated with our default setup.
  Table~\ref{ta:totcs_def} shows integrated LO cross sections, together
  with the (relative) EW corrections, contributions from
  $\gamma\gamma$, gluon fusion and QCD corrections. Moreover, relative
  corrections due to massive-boson radiation (denoted as
  $\delta_{\mathrm{WW}V}$) are included, which will be discussed in the
  following chapter.  In addition, the corresponding results for the
  total cross sections without any event-selection cuts and without a
  dedicated jet veto are shown in Table~\ref{ta:totcs_nocuts} as a
  benchmark.  While our default cuts reduce the LO $q\bar{q}$-induced
  cross section by 44\% (34\%) at the LHC14 (LHC8), the Tevatron cross
  section is only marginally affected. The relative EW corrections are
  changed significantly but remain small as expected. In contrast, the
  relative QCD corrections to the total cross sections are reduced from
  $\sim$ 40\% (30\%) at the LHC (Tevatron) to almost zero (10\%) after
  introducing the jet veto proposed at the end of
  Section~\ref{se:SMinput}. 
\begin{table}
\begin{equation}
\begin{array}{|c||c|c|c|c|c|c|}
\hline
\mbox{default cuts} & \sigma_\LO^{q\bar{q}}\,(\pba)  &
\delta_{\EW}\,(\%) &\delta_{\gamma\gamma} (\%)&
\delta_{\mathrm{gg}} (\%)&
\delta_{\QCD}^{\mathrm{veto}}\,(\%)&\delta_{\mathrm{WW}V}(\%) \nonumber\\
\hline\hline
\mathrm{LHC8}    & 23.99  & - 0.7 & 1.7 & 5.8 &  2.0 & 0.3 \nonumber\\
\mathrm{LHC14}   & 42.39  & - 0.9 & 1.7 & 9.6 &  0.5 & 0.4 \nonumber\\
\hline
\mathrm{Tevatron}&  7.054 & - 0.5 & 0.5 & 0.5 & 11.4 & 0.1 \nonumber\\
\hline
\end{array}
\end{equation}
\caption{\label{ta:totcs_def} Integrated leading-order cross sections
  and relative corrections for the LHC and the Tevatron evaluated with the
  default setup defined in Section~\ref{se:SMinput}.}
\end{table}
\begin{table}
\begin{equation}
\begin{array}{|c||c|c|c|c|c|c|}
\hline
\mbox{no cuts} & \sigma_\LO^{q\bar{q}}\,(\pba)  &
\delta_{\EW}\,(\%) &\delta_{\gamma\gamma} (\%)&
\delta_{\mathrm{gg}} (\%)&
\delta_{\QCD}^{\mathrm{full}}\,(\%)&\delta_{\mathrm{WW}V}(\%) \nonumber\\
\hline\hline
\mathrm{LHC8}    & 35.51  & - 0.4 & 1.5 & 5.4 & 39.4 & 0.3 \nonumber\\
\mathrm{LHC14}   & 75.02  & - 0.4 & 1.6 & 8.1 & 44.2 & 0.3 \nonumber\\
\hline
\mathrm{Tevatron}&  7.916 & - 0.2 & 0.5 & 0.5 & 31.9 & 0.1 \nonumber\\
\hline
\end{array}
\end{equation}
\caption{\label{ta:totcs_nocuts} Total leading-order cross sections and corresponding
  relative corrections for the LHC and the Tevatron without any
  phase-space cuts.}
\end{table}

Predictions for the production cross section with cuts on the transverse
momenta or the invariant mass of the W-pair have been presented already
in Section~\ref{se:details}. For cross checks and to simplify numerical
comparisons we show in Tables~\ref{ta:pt_os} and~\ref{ta:mww_os} the
cross sections and the corresponding relative corrections for different
cut values on the W transverse momenta (Table~\ref{ta:pt_os}) and on the
invariant mass of the W-boson pair (Table~\ref{ta:mww_os}).  While the
relative corrections $\delta_{\mathrm{gg}}$ due to process~\eqref{details:gg} are
negligible at high invariant masses and transverse momenta, we recall
that the photon-induced channels lead to surprisingly large relative
contributions of $\sim +20\%$ at $M_{\PW\PW} \sim 2\;\TeV$, compensating
a significant part of the moderate negative Sudakov-enhanced genuine EW
contributions. This behaviour may be understood qualitatively recalling
that the partonic cross section $\hat{\sigma}_{\gamma\gamma}$ is, for
sufficiently high energies, given by a constant,
$\hat{\sigma}_{\gamma\gamma} = 8\pi\alpha^2/\MW^2 +
\mathcal{O}(1/\hat{s})$, and strongly peaked at small angles, while the
corresponding cross section $\hat{\sigma}_{q\bar{q}}$ for large
$\hat{s}$ decreases as $\ln(\hat{s}/\MW^2)/\hat{s}$.
 Although the NLO QCD corrections turn out to be
accidentally small in our default setup, they still amount to $20\%$ for
large $M_{\PW\PW}^{\mathrm{cut}}$ and $\ptcut$, respectively, exhibiting
a similar behaviour at the LHC and the Tevatron (see
Figs.~\ref{fi:totcs_ptw}, \ref{fi:totcs_mww} and Tables~\ref{ta:pt_os},
\ref{ta:mww_os}).
\begin{table}[]
\begin{equation}
\begin{array}{|c||c|c|c|c|c|c|}
\hline
\ptcut\,(\GeV) & \sigma_\LO^{q\bar{q}}\,(\pba)  &
\delta_{\EW}\,(\%) &\delta_{\gamma\gamma} (\%)&
\delta_{\mathrm{gg}} (\%)&
\delta_{\QCD}^{\mathrm{veto}}\,(\%)&\delta_{\mathrm{WW}V}(\%) \nonumber\\
\hline\hline
%0    & 46.62              & -  0.6 &  1.6  & ??? & ??? \nonumber\\
 50  & 19.80              & -  2.9 &  2.5  & 9.6 &  0.8 & 0.6 \nonumber\\
100  &  5.379             & -  7.0 &  3.8  & 9.2 &  5.3 & 1.0 \nonumber\\
250  & 35.31 \cdot 10^{-2} & - 18.8 &  5.8  & 7.3 & 13.4 & 2.3 \nonumber\\
500  & 23.05 \cdot 10^{-3} & - 33.7 &  7.0  & 3.7 & 15.9 & 4.0 \nonumber\\
750  & 33.04 \cdot 10^{-4} & - 44.9 &  7.9  & 2.2 & 16.9 & 5.3 \nonumber\\
1000 & 67.04 \cdot 10^{-5} & - 53.8 &  8.9  & 1.4 & 17.9 & 6.1 \nonumber\\
1250 & 16.39 \cdot 10^{-5} & - 61.3 & 10.2  & 1.0 & 18.7 & 6.6 \nonumber\\
1500 & 44.79 \cdot 10^{-6} & - 67.7 & 11.8  & 0.8 & 19.7 & 6.9 \nonumber\\
\hline
\end{array}
\end{equation}
\caption{\label{ta:pt_os} Integrated leading-order cross sections and relative  corrections
  for different values of $\ptcut$ at the LHC at
  $\sqrt{s} = 14\,\TeV$, evaluated with the default setup defined in
  Section~\ref{se:SMinput}.}

\end{table}
\begin{table}[]
\begin{equation}
\begin{array}{|c||c|c|c|c|c|c|}
\hline
M_{\mathrm{WW}}^{\mathrm{cut}}\,(\GeV) & \sigma_\LO^{q\bar{q}}\,(\pba) & 
\delta_{\EW}\,(\%) & \delta_{\gamma\gamma} (\%)&
\delta_{\mathrm{gg}} (\%)&
\delta_{\QCD}^{\mathrm{veto}}\,(\%)&\delta_{\mathrm{WW}V}(\%) \nonumber\\
\hline\hline
200  & 28.84              & - 2.1 &  2.2 & 8.9 &  7.1 & 0.5 \nonumber\\
300  & 9.492              & - 4.0 &  3.8 & 6.4 & 13.5 & 0.8 \nonumber\\
500  & 1.841              & - 7.5 &  7.2 & 4.8 & 18.8 & 1.4 \nonumber\\
1000 & 12.08 \cdot 10^{-2} & -17.6 & 14.4 & 2.6 & 24.1 & 3.2 \nonumber\\
1500 & 20.37 \cdot 10^{-3} & -25.4 & 18.1 & 1.4 & 23.1 & 4.2 \nonumber\\
2000 & 48.79 \cdot 10^{-4} & -31.3 & 21.6 & 0.9 & 22.8 & 4.9 \nonumber\\
2500 & 13.81 \cdot 10^{-4} & -36.2 & 25.6 & 0.6 & 22.5 & 5.2 \nonumber\\
3000 & 42.99 \cdot 10^{-5} & -40.5 & 30.5 & 0.4 & 22.3 & 5.4 \nonumber\\
\hline
\end{array}
\end{equation}
\caption{\label{ta:mww_os} Integrated leading-order cross sections and relative corrections for different values of
  $M_{\mathrm{WW}}^{\mathrm{cut}}$ at the LHC at $\sqrt{s} = 14\,\TeV$,
  evaluated with the default setup defined in
  Section~\ref{se:SMinput}.}
\end{table}

\subsection{Transverse-momentum, invariant-mass and rapidity
  distributions}
\begin{figure}[]
\begin{center}
\includegraphics[width = 1.0\textwidth]{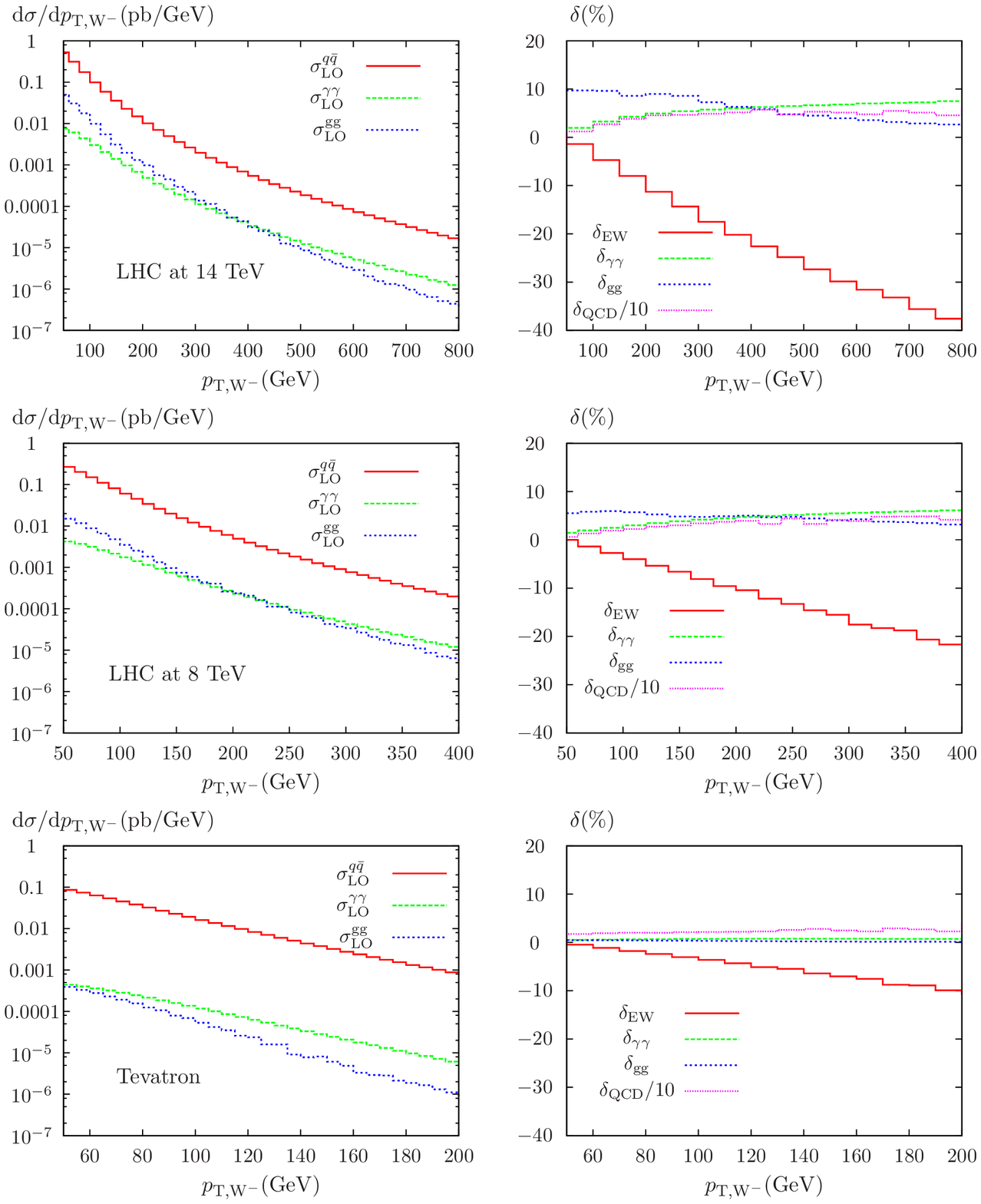} 
\end{center}
\caption{\label{fi:histptwm} Distributions of the $\mathrm{W}^-$ transverse
  momentum at the LHC14 (top), LHC8 (center) and the Tevatron
  (bottom). On the left-hand side, LO contributions due to
  processes~\eqref{details:LO}($q\bar{q}$),
  ~\eqref{details:gaga}($\gamma\gamma$), and~\eqref{details:gg}($\mathrm{gg}$)
  are shown. On the right-hand side, corresponding relative corrections are
  presented, normalized to the dominating LO
  channel~\eqref{details:LO}. See text for details.}
\end{figure}
\begin{figure}[]
\begin{center}
\includegraphics[width = 1.0\textwidth]{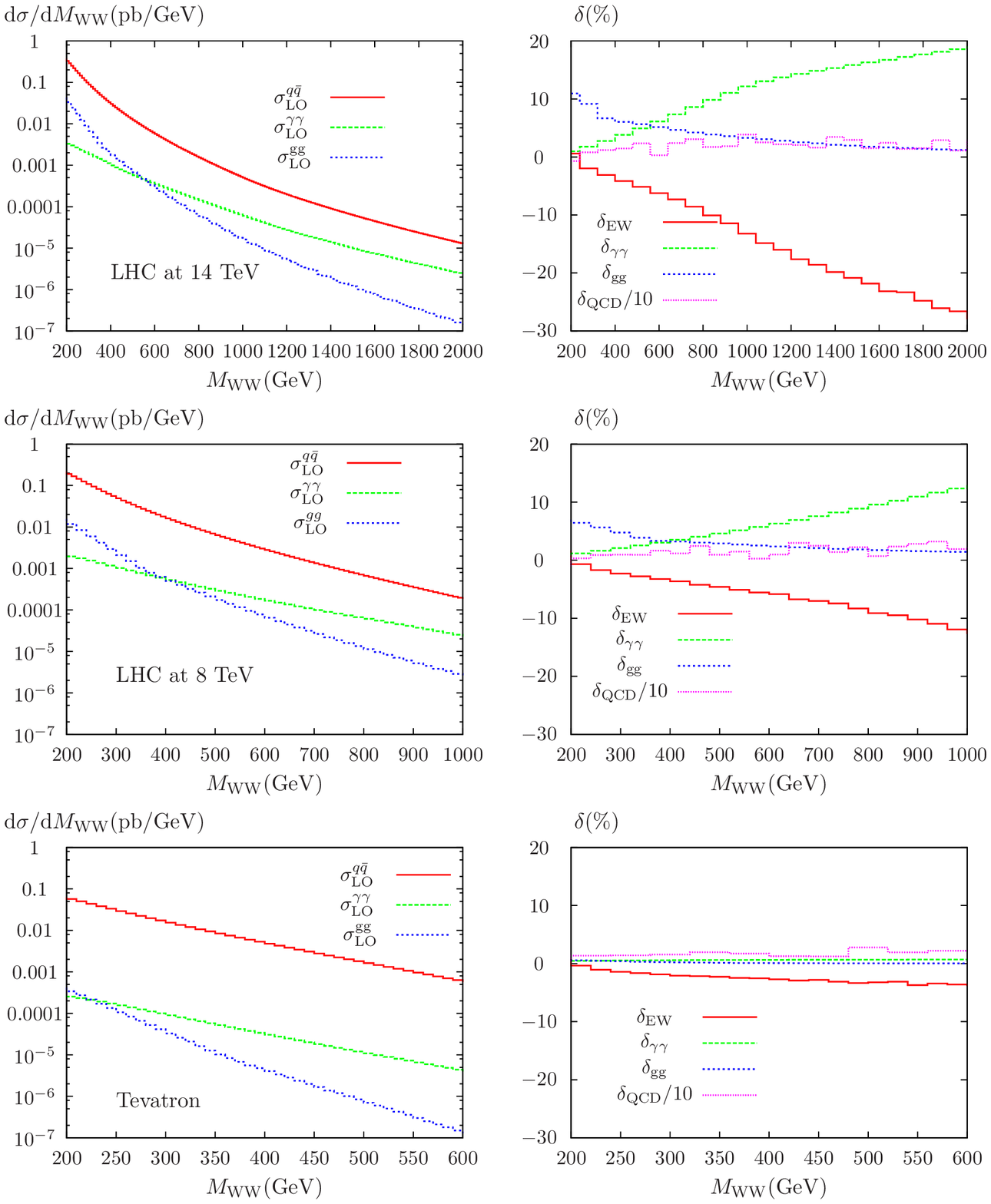} 
\end{center}
\caption{\label{fi:histmww} Distributions of the invariant mass of the
  W-boson pair at the LHC14 (top), LHC8 (center) and the Tevatron
  (bottom). On the left-hand side, LO contributions due to
  processes~\eqref{details:LO}($q\bar{q}$),
  ~\eqref{details:gaga}($\gamma\gamma$), and~\eqref{details:gg}($\mathrm{gg}$)
  are shown. On the right-hand side, corresponding relative
  corrections are presented, normalized to the dominating LO
  channel~\eqref{details:LO}. See text for details.}
\end{figure}

\begin{figure}[]
\begin{center}
\includegraphics[width = 1.0\textwidth]{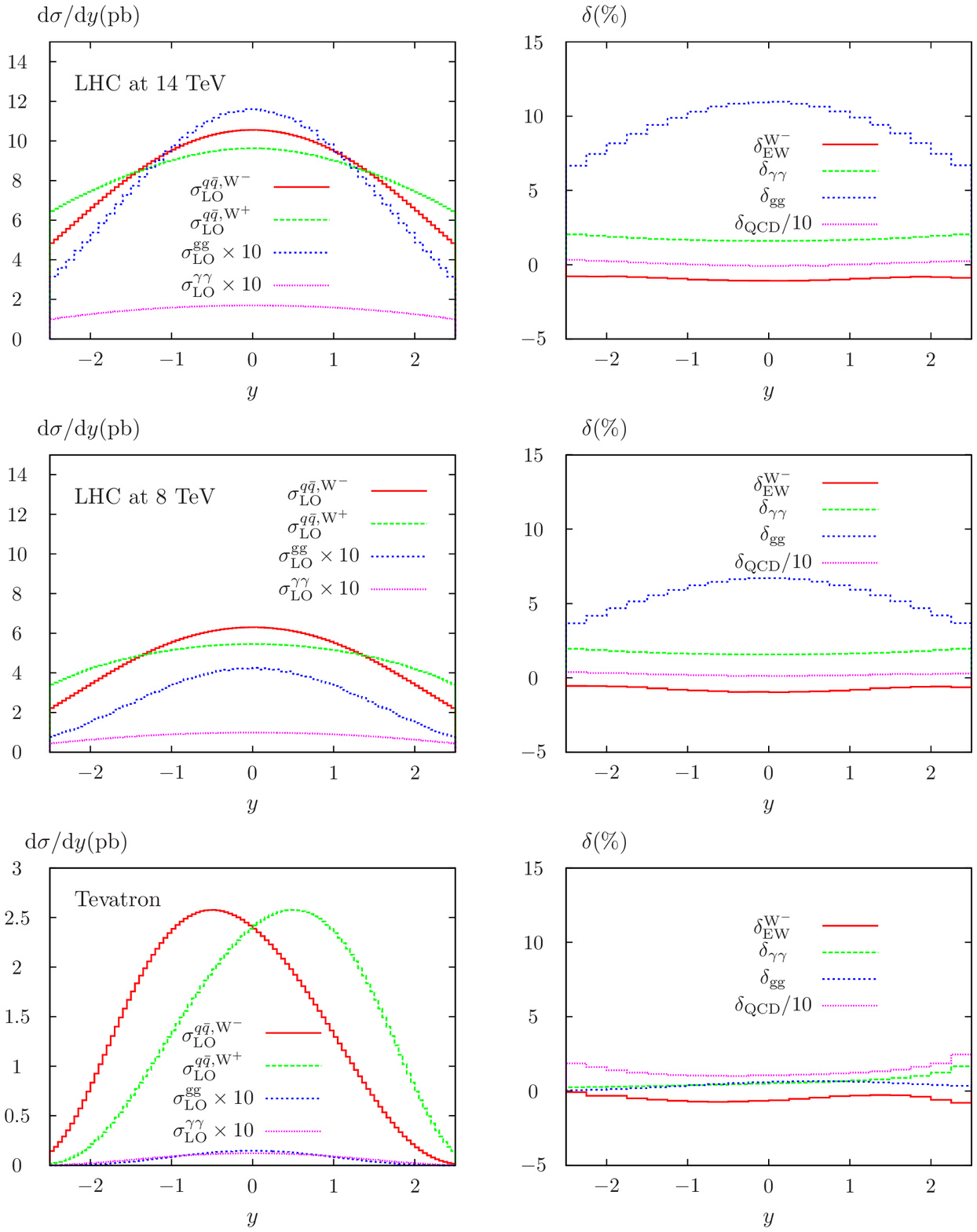} 
\end{center}
\caption{\label{fi:histyw} Distributions of the rapidities of the
  W-bosons at the LHC14 (top), LHC8 (center) and the Tevatron
  (bottom). On the left-hand side, LO contributions due to
  processes~\eqref{details:LO}($q\bar{q}$),
  ~\eqref{details:gaga}($\gamma\gamma$), and~\eqref{details:gg}($\mathrm{gg}$)
  are shown. On the right-hand side, corresponding relative
  corrections are presented, normalized to the dominating LO
  channel~\eqref{details:LO}. See text for details.}
\end{figure}
\begin{figure}[]
\begin{center}
\includegraphics[width = 1.0\textwidth]{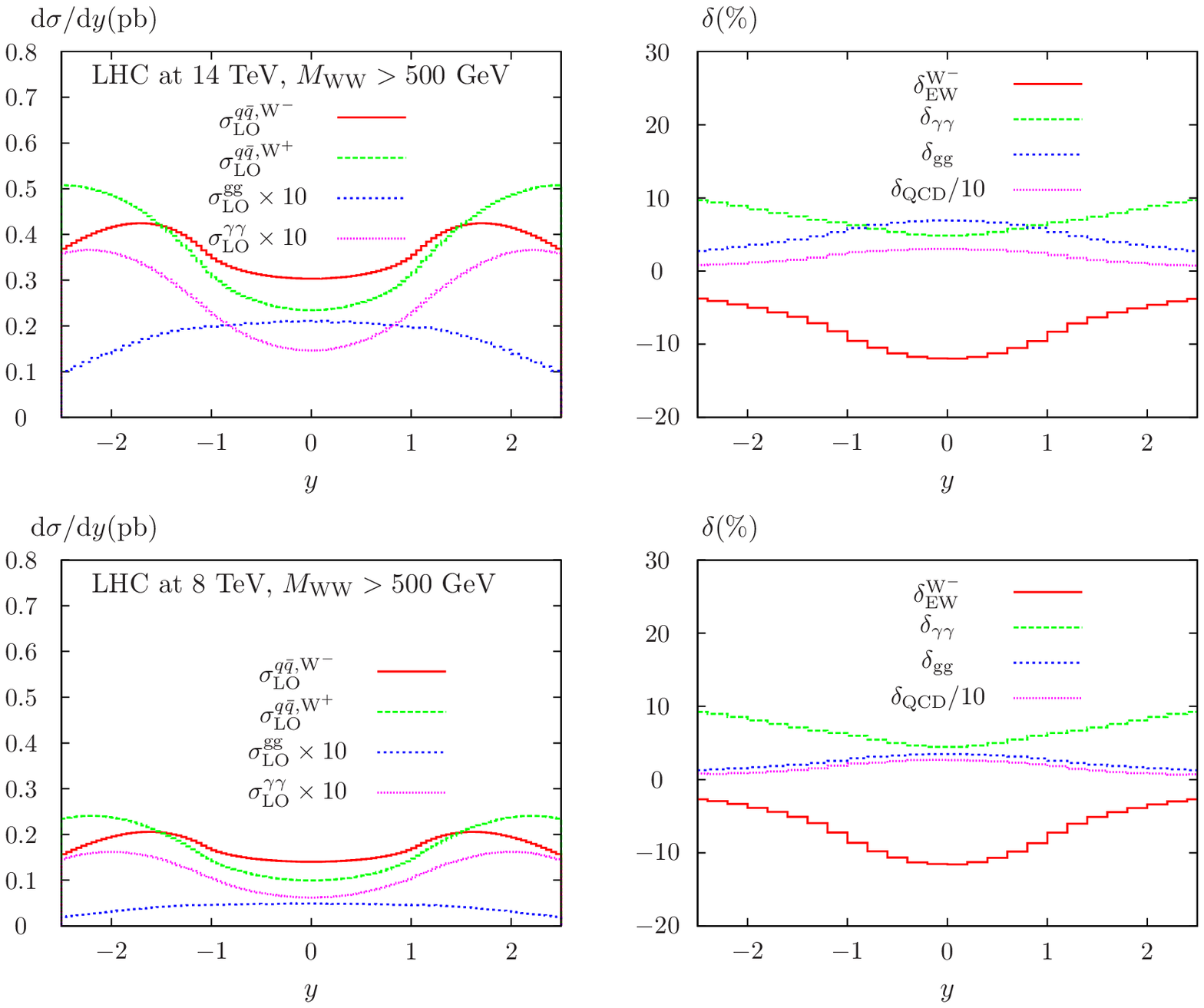}
\end{center}
\caption{\label{fi:histyw500} Distributions of the rapidities of the
  W-bosons at the LHC14 (top) and LHC8 (bottom) for 
  $M_{\PW\PW}>500\;\GeV$. On the left-hand side, LO contributions due to
  processes~\eqref{details:LO}($q\bar{q}$),
  ~\eqref{details:gaga}($\gamma\gamma$), and~\eqref{details:gg}($\mathrm{gg}$)
  are shown. On the right-hand side, corresponding relative corrections
  are presented, normalized to the dominating LO
  channel~\eqref{details:LO}. See text for details.}
\end{figure}
\begin{figure}[]
\begin{center}
\includegraphics[width = 1.0\textwidth]{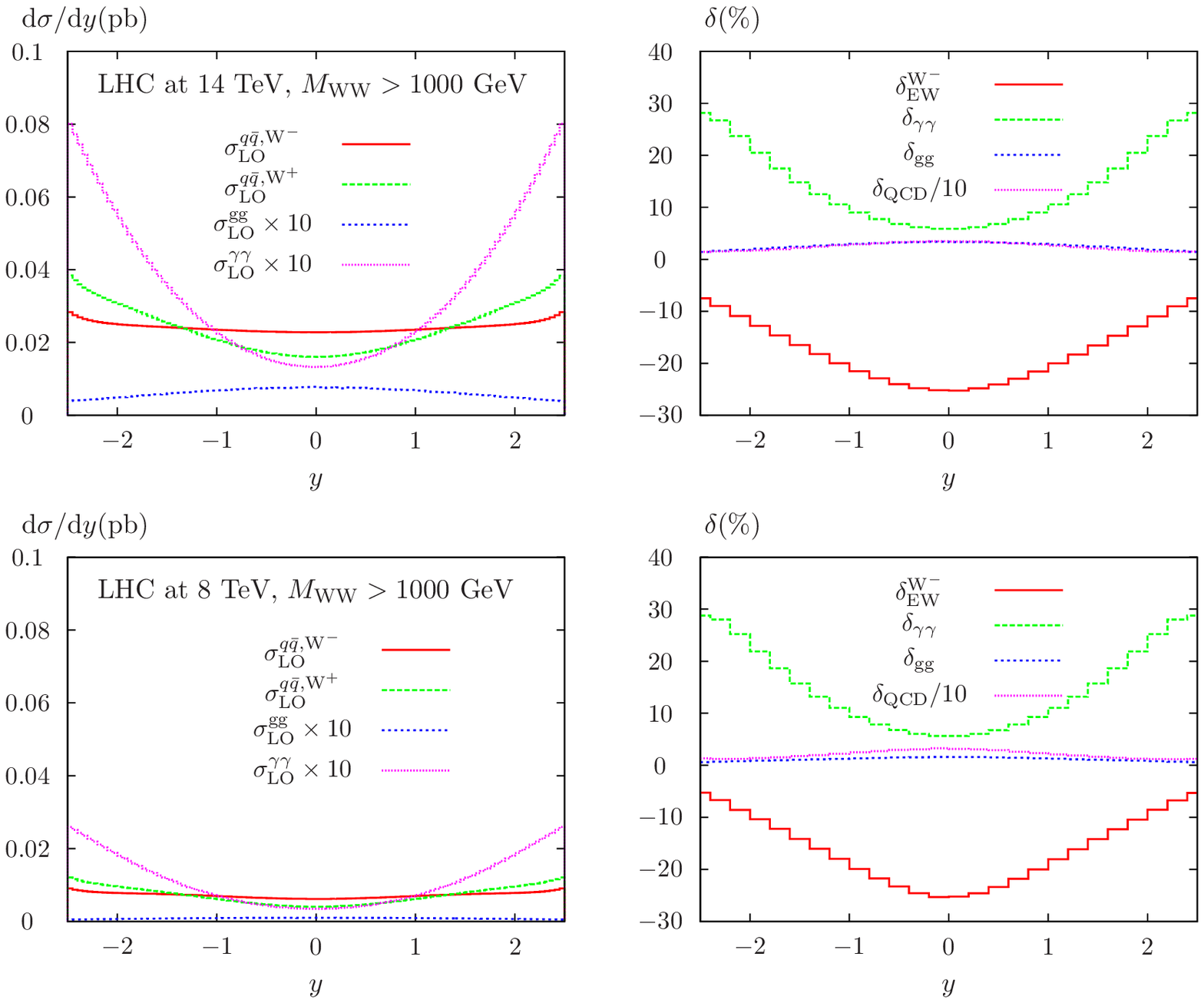} 
\end{center}
\caption{\label{fi:histyw1000} Distributions of the rapidities of the
  W-bosons at the LHC14 (top) and LHC8 (bottom) for
  $M_{\PW\PW}>1000\;\GeV$. On 
  the left-hand side, LO contributions due to
  processes~\eqref{details:LO}($q\bar{q}$),
  ~\eqref{details:gaga}($\gamma\gamma$), and~\eqref{details:gg}($\mathrm{gg}$)
  are shown. On the right-hand side, corresponding relative
  corrections are presented, normalized to the dominating LO
  channel~\eqref{details:LO}. See text for details.}
\end{figure}

\begin{figure}[]
\begin{center}
\includegraphics[width = 1.0\textwidth]{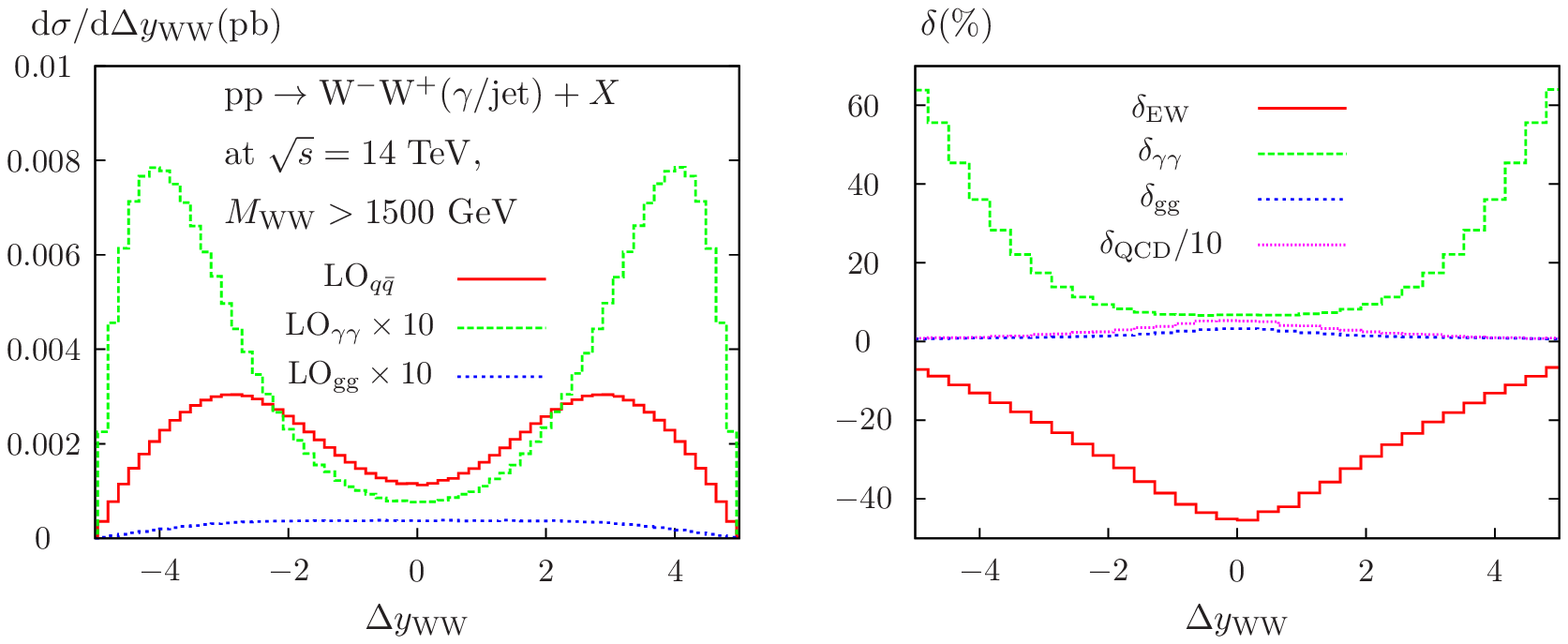} 
\end{center}
\caption{\label{fi:histdyww1500} Distributions of the W-boson rapidity
  gap at the LHC14 for $M_{\PW\PW}>1500\;\GeV$. On
  the left-hand side, LO contributions due to
  processes~\eqref{details:LO}($q\bar{q}$),
  ~\eqref{details:gaga}($\gamma\gamma$), and~\eqref{details:gg}($\mathrm{gg}$)
  are shown. On the right-hand side, corresponding relative
  corrections are presented, normalized to the dominating LO
  channel~\eqref{details:LO}. See text for details.}
\end{figure}

\begin{figure}[]
\begin{center}
\includegraphics[width = 1.0\textwidth]{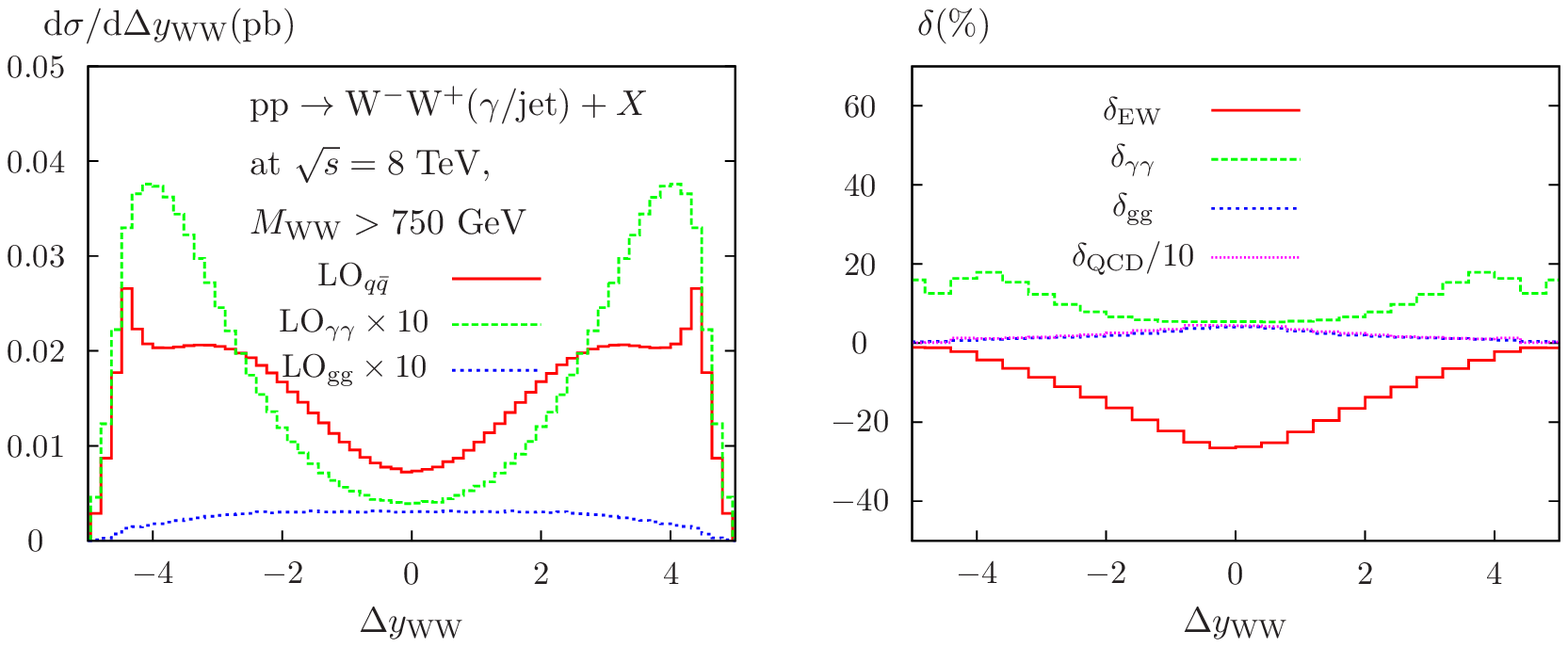} 
\end{center}
\caption{\label{fi:histdyww750} Distributions of the W-boson rapidity
  gap at the LHC8 for $M_{\PW\PW}>750\;\GeV$. On
  the left-hand side, LO contributions due to
  processes~\eqref{details:LO}($q\bar{q}$),
  ~\eqref{details:gaga}($\gamma\gamma$), and~\eqref{details:gg}($\mathrm{gg}$)
  are shown. On the right-hand side, corresponding relative
  corrections are presented, normalized to the dominating LO
  channel~\eqref{details:LO}. See text for details.}
\end{figure}
In a first step we show the transverse-momentum distributions of
$\mathrm{W}^-$ and $\mathrm{W}^+$ (Fig.~\ref{fi:histptwm}) and the
distributions of the invariant mass $M_{\PW\PW}$ (Fig.~\ref{fi:histmww})
for LHC14, LHC8 and the Tevatron. The interpretation of the results is
similar to that of the partially integrated cross sections presented in
Figs.~\ref{fi:totcs_ptw} and~\ref{fi:totcs_mww} in
Section~\ref{se:details}. In particular, we again observe the large
Sudakov logarithms resulting in corrections of $-30\%$ for $\pT$ values
around 800 GeV. Again, the correction for the invariant-mass
distribution is smaller, since large $M_{\PW\PW}$ may still involve
small momentum transfer $\hat{t}$. At the LHC, the QCD corrections
evaluated with the jet veto proposed in Section~\ref{se:SMinput} still
reach 50\% (20\%) at high $p_{\rT}$ ($M_{\PW\PW}$), while at the
Tevatron they do not exceed 20\% even at large transverse momenta.

Rapidity distributions of $\mathrm{W}^+$ and $\mathrm{W}^-$ individually
are shown in Fig.~\ref{fi:histyw}. Note that the rapidity distributions
of $\mathrm{W}^+$ and $\mathrm{W}^-$ are different at the LHC, as a
consequence of the asymmetric piece of the differential cross section
for \mbox{$\mathrm{u}\bar{\mathrm{u}} \to \mathrm{W^-W^+}$} and
\mbox{$\mathrm{d}\bar{\mathrm{d}} \to \mathrm{W^-W^+}$}, and the
difference between valence- and sea-quark distributions. For the
Tevatron one finds \mbox{$\rd\sigma_{\PW^+}(y) =
  \rd\sigma_{\PW^-}(-y)$}.  At the LHC8 (LHC14), gluon fusion and the
$\gamma\gamma$ process increase the W-pair production rate by $\sim 5\%$
(10\%) and $\sim 2\%$ (2\%), respectively. The EW corrections are $\sim
-1\%$, reflecting the small $\hat{s}$ of typical events and the absence
of the large negative Sudakov logarithms. At the
Tevatron, only QCD corrections must be considered for this particular
observable; EW corrections, \mbox{$\gamma\gamma$ and $\mathrm{gg}$
  fusion} can safely be neglected.

The rapidity distributions for the LHC14 and the LHC8 for the
subsamples with invariant mass \mbox{$M_{\PW\PW}>500\;\GeV$} and
\mbox{$M_{\PW\PW}>1000\;\GeV$} 
are shown in Figs.~\ref{fi:histyw500} and~\ref{fi:histyw1000},
respectively.  In these cases EW corrections and the $\gamma\gamma$
process are significantly more important. 

Let us finally consider W-pair production at the highest energies
accessible at the LHC. As already mentioned in Section~\ref{se:details},
the distribution in the rapidity difference (for fixed $M_{\PW\PW}$)
between $\mathrm{W}^+$ and $\mathrm{W}^-$ bosons can be directly related
to the angular distribution in the $\mathrm{WW}$ rest frame and thus can
be used to search for anomalous couplings at the TeV scale. In
Figs.~\ref{fi:histdyww1500} and~\ref{fi:histdyww750} we therefore show
these distributions for $M_{\PW\PW} > 1.5 \;\TeV$ for LHC14 and
$M_{\PW\PW} > 0.75 \;\TeV $ for LHC8, respectively. At the LHC14, large
contributions are visible both from the $\gamma\gamma$ process
($+60\%$) and from the weak corrections ($-45\%$),
leading to a strong distortion of this
particular distribution. As already discussed in
  Section~\ref{se:details}, 
the additional contributions from $\mathrm{gg}$ and $\gamma\gamma$
fusion are small in the region accessible to the Tevatron and hence will
not be discussed further.

Finally, we compare our predictions to older results obtained in the
high-energy approximation.  Unfortunately, a tuned comparison of our
results with the ones presented in Ref.~\cite{Accomando:2004de} is not
possible, since in that paper dedicated event-selection cuts on the
leptonic decay products of the W bosons are applied. However, comparing
Fig.~7 (bottom, left) from Ref.~\cite{Accomando:2004de} with
Fig.~\ref{fi:dy_LO} (bottom, right), we find a reasonable agreement for
the relative EW corrections within a few percent. Here, we assume that
each of the charged leptons on average carries away 50\% of the momentum
of the decaying W, and a strong correlation between $\Delta y_{\PW\PW}$
and $\Delta y _{l\bar{l'}}$, which seems to be justified in case of
strongly-boosted Ws.

\section{Radiation of massive vector bosons}
\label{se:real}
\begin{figure}[]
\begin{center}
\includegraphics[width = 1.0\textwidth]{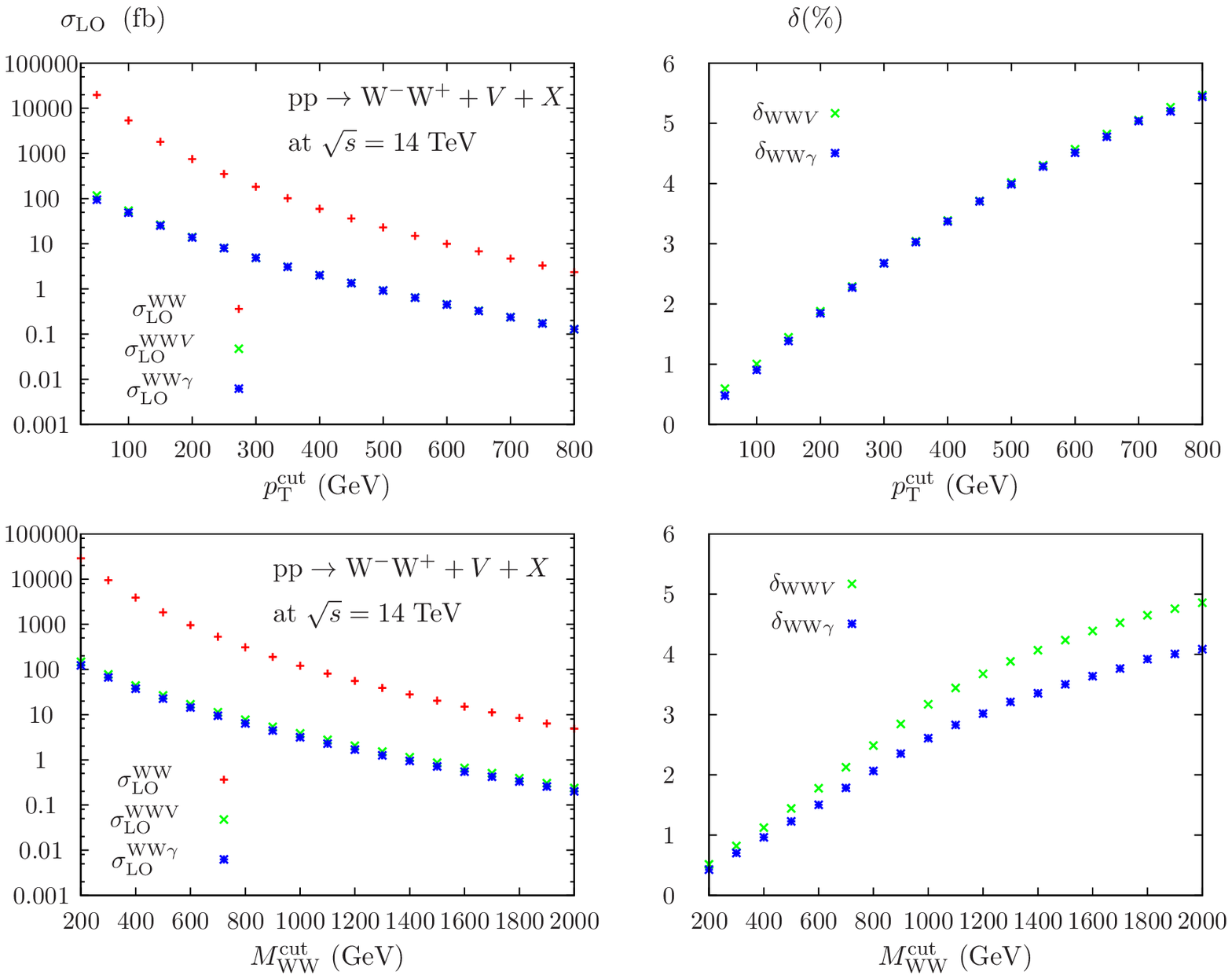} 
\end{center}
\caption{\label{fi:WWV} Left: Total LO cross sections for $\mathrm{WW}$,
  $\mathrm{WW}V$ ($V=\mathrm{W},\mathrm{Z}$) and $\mathrm{WW}\gamma$
  production; Right: Corrections due to $\mathrm{WW}V$ and $\mathrm{WW}
  \gamma$ relative to quark-induced W-pair production. See text for
    details.}
\end{figure}
Finally, we discuss the phenomenological effects of additional massive
vector bosons in the final state produced in the partonic processes
\begin{eqnarray}
  {\rm u}_i\mathrm{\bar{d}}_j &\to& \mathrm{W}^-\mathrm{W}^+\mathrm{W}^+\,,\\
  \mathrm{\bar{u}}_i\mathrm{d}_j &\to& \mathrm{W}^-\mathrm{W}^+\mathrm{W}^-\,,\\
  q\bar{q} &\to& \mathrm{W}^-\mathrm{W}^+\mathrm{Z}\,. 
\end{eqnarray}
Although these channels are parametrically suppressed by one order of
the EW coupling, collinear or soft massive boson radiation may
potentially lead to logarithmically enhanced contributions and thus
needs further investigation.  In the numerical analysis we use a
simplified approach to conservatively estimate the pollution of W-pair
production which can be expected from \mbox{3-boson} final states; we
treat the additional boson completely inclusively, whereas our default
cuts are applied to those opposite-sign W-bosons with highest
$p_{\rT}$. In Fig.~\ref{fi:WWV} we present the corresponding numerical
results for the $\mathrm{WW}V$ ($V=\mathrm{W},\mathrm{Z}$) final state
as well as the contribution from hard-photon radiation
($p_{\rT,\gamma}>15$ GeV, $|y_\gamma|<2.5$) relative to quark-induced
$\mathrm{WW}$ production at leading order. (The corresponding numbers
can be found in Tables~\ref{ta:pt_os} and~\ref{ta:mww_os},
respectively.)  
The relative corrections due to massive boson radiation
are below 5\% even for large transverse momenta and invariant masses and
therefore of minor importance.  Remarkably enough the contribution from
hard photon radiation with these cuts (which is included in the cross
sections discussed in the previous chapter) is numerically close
to the one from additional massive boson radiation.

\section{Conclusions}
\label{se:concl}
We have calculated the full NLO EW corrections to \PW-boson pair
production at hadron colliders and present predictions valid in the full
energy reach of the LHC. At large parton CM energies, the relative corrections
are dominated by the well-known universal Sudakov logarithms that
lead to substantial negative contributions, while the QCD corrections
turn out to be moderate after application of a dynamic jet veto.

As a surprising new result we find that photon-induced contributions can
be of the same size as the genuine EW corrections at high energies and
moderate scattering angles and thus must not be neglected when
predicting W-boson pair production at highest energies. At low
scattering angles, however, the $\gamma\gamma$ channel dominates the EW
contributions even at moderate energies. In the future, the leptonic
decays of the W bosons should be included properly, allowing for a
realistic event definition to match the increasing accuracy of future
LHC measurements.

\subsection*{Acknowledgements}
We would like to thank Peter Marquard for fruitful discussions. This work
has been supported by ``Strukturiertes Promotionskolleg
Elementarteilchen- und Astroteilchenphysik'', SFB TR9 ``Computational
and Particle Physics'' and BMBF Contract 05HT4VKATI3.

\end{document}

%% file: setup_jhep.tex
\newcommand{\GeV}{\mathrm{GeV}}

\newcommand{\TeV}{\mathrm{TeV}}

\newcommand{\pba}{\mathrm{pb}}

\def\mathswitch#1{\relax\ifmmode#1\else$#1$\fi}
\def\mathswitchr#1{\relax\ifmmode{\mathrm{#1}}\else$\mathrm{#1}$\fi}
\def\mathswitchit#1{\relax\ifmmode{#1}\else$#1$\fi}
\newcommand{\muF}{\mu_{\mathrm{F}}}
\newcommand{\muR}{\mu_{\mathrm{R}}}

\newcommand{\PW}{\mathswitchr W}
\newcommand{\PZ}{\mathswitchr Z}

\newcommand{\PH}{\mathswitchr H}

\newcommand{\Pu}{\mathrm{u}}
\newcommand{\Pub}{\mathrm{\bar{u}}}

\newcommand{\Pb}{\mathrm{b}}
\newcommand{\Pbb}{\mathrm{\bar{b}}}
\newcommand{\Pt}{\mathrm{t}}

\newcommand{\MW}{\mathswitch {M_\PW}}

\newcommand{\MZ}{\mathswitch {M_\PZ}}

\newcommand{\Mt}{\mathswitch {m_\Pt}}

\newcommand{\GF}{\mathswitch {G_\mu}}

\newcommand{\WW}{\mathrm{WW}}
\newcommand{\ptcut}{p_{\mathrm{T}}^{\mathrm{cut}}}
\newcommand{\alphas}{\alpha_{\mathrm{s}}}
\newcommand{\rT}{{\mathrm{T}}}

\newcommand{\rd}{{\mathrm{d}}}

\newcommand{\M}{{\cal {M}}}

\newcommand{\pT}{p_\mathrm{T}}
\newcommand{\De}{\Delta}
\newcommand{\al}{\alpha}

\newcommand{\jet}{{\mathrm{jet}}}
\newcommand{\cut}{{\mathrm{cut}}}

\newcommand{\EW}{{\mathrm{EW}}}
\newcommand{\QCD}{{\mathrm{QCD}}}

\newcommand{\LO}{{\mathrm{LO}}}
\newcommand{\NLO}{{\mathrm{NLO}}}

\def\Re{\mathop{\mathrm{Re}}\nolimits}

\def\draftdate{\relax}
\def\mda{\relax}
\def\mua{\relax}
\def\mla{\relax}
\def\draft{
\def\thtystars{******************************}
\def\sixtystars{\thtystars\thtystars}
\typeout{}
\typeout{\sixtystars**}
\typeout{* Draft mode!
         For final version remove \protect\draft\space in source file *}
\typeout{\sixtystars**}
\typeout{}
\def\draftdate{\today}
\def\mua{\marginpar[\boldmath\hfil$\uparrow$]%
                   {\boldmath$\uparrow$\hfil}%
                    \typeout{marginpar: $\uparrow$}\ignorespaces}
\def\mda{\marginpar[\boldmath\hfil$\downarrow$]%
                   {\boldmath$\downarrow$\hfil}%
                    \typeout{marginpar: $\downarrow$}\ignorespaces}
\def\mla{\marginpar[\boldmath\hfil$\rightarrow$]%
                   {\boldmath$\leftarrow $\hfil}%
                    \typeout{marginpar: $\leftrightarrow$}\ignorespaces}
\def\Mua{\marginpar[\boldmath\hfil$\Uparrow$]%
                   {\boldmath$\Uparrow$\hfil}%
                    \typeout{marginpar: $\Uparrow$}\ignorespaces}
\def\Mda{\marginpar[\boldmath\hfil$\Downarrow$]%
                   {\boldmath$\Downarrow$\hfil}%
                    \typeout{marginpar: $\Downarrow$}\ignorespaces}
\def\Mla{\marginpar[\boldmath\hfil$\Rightarrow$]%
                   {\boldmath$\Leftarrow $\hfil}%
                    \typeout{marginpar: $\Leftrightarrow$}\ignorespaces}
\overfullrule 5pt
\oddsidemargin -15mm
\marginparwidth 29mm
}

\makeatother

%% file: diags/diaguuWW.tex
\unitlength=2.3bp%
%\SetScale{2.bp}
\begin{center}
\begin{small}
\begin{feynartspicture}(120,40)(3,1)
%\FALabel(33.,69.96)[]{a)}
\FADiagram{}
\FAProp(0.,15.)(6.,10.)(0.,){/Straight}{1}
\FALabel(2.48771,11.7893)[tr]{$\mathrm{u}$}
\FAProp(0.,5.)(6.,10.)(0.,){/Straight}{-1}
\FALabel(3.51229,6.78926)[tl]{$\mathrm{\bar{u}}$}
\FAProp(20.,15.)(14.,10.)(0.,){/Sine}{-1}
\FALabel(16.4877,13.2107)[br]{$\mathrm{W}$}
\FAProp(20.,5.)(14.,10.)(0.,){/Sine}{1}
\FALabel(17.5123,8.21074)[bl]{$\mathrm{W}$}
\FAProp(6.,10.)(14.,10.)(0.,){/Sine}{0}
\FALabel(10.,8.93)[t]{$\mathrm{Z}/\gamma$}
\FAVert(6.,10.){0}
\FAVert(14.,10.){0}

\FADiagram{}
\FAProp(0.,15.)(10.,14.)(0.,){/Straight}{1}
\FALabel(4.84577,13.4377)[t]{$\mathrm{u}$}
\FAProp(0.,5.)(10.,6.)(0.,){/Straight}{-1}
\FALabel(5.15423,4.43769)[t]{$\mathrm{\bar{u}}$}
\FAProp(20.,15.)(10.,14.)(0.,){/Sine}{-1}
\FALabel(14.8458,15.5623)[b]{$\mathrm{W}$}
\FAProp(20.,5.)(10.,6.)(0.,){/Sine}{1}
\FALabel(15.1542,6.56231)[b]{$\mathrm{W}$}
\FAProp(10.,14.)(10.,6.)(0.,){/Straight}{1}
\FALabel(9.18,10.)[r]{$\mathrm{d}$}
\FAVert(10.,14.){0}
\FAVert(10.,6.){0}

\FADiagram{}
\FAProp(0.,15.)(10.,14.)(0.,){/Straight}{1}
\FALabel(4.84577,13.4377)[t]{$\mathrm{u}$}
\FAProp(0.,5.)(10.,6.)(0.,){/Straight}{-1}
\FALabel(5.15423,4.43769)[t]{$\mathrm{\bar{u}}$}
\FAProp(20.,15.)(10.,6.)(0.,){/Sine}{-1}
\FALabel(16.8128,13.2058)[br]{$\mathrm{W}$}
\FAProp(20.,5.)(10.,14.)(0.,){/Sine}{1}
\FALabel(17.6872,8.20582)[bl]{$\mathrm{W}$}
\FAProp(10.,14.)(10.,6.)(0.,){/Straight}{1}
\FALabel(9.28,10.)[r]{$\mathrm{d}$}
\FAVert(10.,14.){0}
\FAVert(10.,6.){0}

\end{feynartspicture}
\end{small}
\end{center}

%% file: diags/diaguuWWphot.tex
\unitlength=2.3bp%

\begin{center}
\begin{small}
\begin{feynartspicture}(120,40)(3,1)
%\FALabel(33.,69.96)[]{b}

\FADiagram{}
\FAProp(0.,15.)(10.,10.)(0.,){/Sine}{0}
\FALabel(4.78682,11.5936)[tr]{$\gamma$}
\FAProp(0.,5.)(10.,10.)(0.,){/Sine}{0}
\FALabel(5.21318,6.59364)[tl]{$\gamma$}
\FAProp(20.,15.)(10.,10.)(0.,){/Sine}{-1}
\FALabel(14.7868,13.4064)[br]{$\mathrm{W}$}
\FAProp(20.,5.)(10.,10.)(0.,){/Sine}{1}
\FALabel(15.2132,8.40636)[bl]{$\mathrm{W}$}
\FAVert(10.,10.){0}

\FADiagram{}
\FAProp(0.,15.)(10.,14.)(0.,){/Sine}{0}
\FALabel(4.84577,13.4377)[t]{$\gamma$}
\FAProp(0.,5.)(10.,6.)(0.,){/Sine}{0}
\FALabel(5.15423,4.43769)[t]{$\gamma$}
\FAProp(20.,15.)(10.,14.)(0.,){/Sine}{-1}
\FALabel(14.8458,15.5623)[b]{$\mathrm{W}$}
\FAProp(20.,5.)(10.,6.)(0.,){/Sine}{1}
\FALabel(15.1542,6.56231)[b]{$\mathrm{W}$}
\FAProp(10.,14.)(10.,6.)(0.,){/Sine}{-1}
\FALabel(8.93,10.)[r]{$\mathrm{W}$}
\FAVert(10.,14.){0}
\FAVert(10.,6.){0}

\FADiagram{}
\FAProp(0.,15.)(10.,14.)(0.,){/Sine}{0}
\FALabel(4.84577,13.4377)[t]{$\gamma$}
\FAProp(0.,5.)(10.,6.)(0.,){/Sine}{0}
\FALabel(5.15423,4.43769)[t]{$\gamma$}
\FAProp(20.,15.)(10.,6.)(0.,){/Sine}{-1}
\FALabel(16.8128,13.2058)[br]{$\mathrm{W}$}
\FAProp(20.,5.)(10.,14.)(0.,){/Sine}{1}
\FALabel(17.6872,8.20582)[bl]{$\mathrm{W}$}
\FAProp(10.,14.)(10.,6.)(0.,){/Sine}{1}
\FALabel(9.03,10.)[r]{$\mathrm{W}$}
\FAVert(10.,14.){0}
\FAVert(10.,6.){0}

\end{feynartspicture}
\end{small}
\end{center}

%% file: diags/diaguuWWreal.tex
\unitlength=2.3bp%
%\SetScale{2.5bp}
\begin{center}
\begin{small}
\begin{feynartspicture}(160,120)(4,3)
%\FALabel(33.,69.96)[]{\large $u\quad u\quad \to\quad W\quad W\quad \gamma$}

\FADiagram{}
\FAProp(0.,15.)(5.5,10.)(0.,){/Straight}{1}
\FALabel(-0.5,16)[tr]{$q$}
%\FALabel(2.18736,11.8331)[tr]{$q$}
\FAProp(0.,5.)(5.5,10.)(0.,){/Straight}{-1}
\FALabel(-0.5,6)[tr]{$\bar q$}
%\FALabel(3.31264,6.83309)[tl]{$\bar q$}
\FAProp(20.,17.)(12.5,10.)(0.,){/Sine}{-1}
\FALabel(19,16)[br]{$\mathrm{W}$}
%\FALabel(15.6724,14.1531)[br]{$\mathrm{W}$}
\FAProp(20.,10.)(12.5,10.)(0.,){/Sine}{1}
\FALabel(17.6,11.07)[b]{$\mathrm{W}$}
\FAProp(20.,3.)(12.5,10.)(0.,){/Sine}{0}
%\FALabel(15.6724,5.84686)[tr]{$\gamma$}
\FALabel(19,7)[tr]{$\gamma$}
\FAProp(5.5,10.)(12.5,10.)(0.,){/Sine}{0}
\FALabel(9.,8.93)[t]{$\mathrm{Z}/\gamma$}
\FAVert(5.5,10.){0}
\FAVert(12.5,10.){0} 

\FADiagram{}
\FAProp(0.,15.)(4.5,10.)(0.,){/Straight}{1}
\FALabel(1.57789,11.9431)[tr]{$q$}
\FAProp(0.,5.)(4.5,10.)(0.,){/Straight}{-1}
\FALabel(2.92211,6.9431)[tl]{$\bar q$}
\FAProp(20.,17.)(13.,14.5)(0.,){/Sine}{-1}
\FALabel(15.9787,16.7297)[b]{$\mathrm{W}$}
\FAProp(20.,10.)(10.85,8.4)(0.,){/Sine}{1}
\FALabel(18.4569,10.7663)[b]{$\mathrm{W}$}
\FAProp(20.,3.)(13.,14.5)(0.,){/Sine}{0}
\FALabel(17.7665,4.80001)[tr]{$\gamma$}
\FAProp(4.5,10.)(10.85,8.4)(0.,){/Sine}{0}
\FALabel(7.29629,8.17698)[t]{$\mathrm{Z/\gamma}$}
\FAProp(13.,14.5)(10.85,8.4)(0.,){/ScalarDash}{-1}
\FALabel(11.1789,11.8821)[r]{$\phi^-$}
\FAVert(4.5,10.){0}
\FAVert(13.,14.5){0}
\FAVert(10.85,8.4){0}

\FADiagram{}
\FAProp(0.,15.)(4.5,10.)(0.,){/Straight}{1}
\FALabel(1.57789,11.9431)[tr]{$q$}
\FAProp(0.,5.)(4.5,10.)(0.,){/Straight}{-1}
\FALabel(2.92211,6.9431)[tl]{$\bar q$}
\FAProp(20.,17.)(13.,14.5)(0.,){/Sine}{-1}
\FALabel(15.9787,16.7297)[b]{$\mathrm{W}$}
\FAProp(20.,10.)(10.85,8.4)(0.,){/Sine}{1}
\FALabel(18.4569,10.7663)[b]{$\mathrm{W}$}
\FAProp(20.,3.)(13.,14.5)(0.,){/Sine}{0}
\FALabel(17.7665,4.80001)[tr]{$\gamma$}
\FAProp(4.5,10.)(10.85,8.4)(0.,){/Sine}{0}
\FALabel(7.29629,8.17698)[t]{$\mathrm{Z}/\gamma$}
\FAProp(13.,14.5)(10.85,8.4)(0.,){/Sine}{-1}
\FALabel(10.9431,11.9652)[r]{$\mathrm{W}$}
\FAVert(4.5,10.){0}
\FAVert(13.,14.5){0}
\FAVert(10.85,8.4){0}

\FADiagram{}
\FAProp(0.,15.)(5.5,10.)(0.,){/Straight}{1}
\FALabel(2.18736,11.8331)[tr]{$q$}
\FAProp(0.,5.)(5.5,10.)(0.,){/Straight}{-1}
\FALabel(3.31264,6.83309)[tl]{$\bar q$}
\FAProp(20.,17.)(11.5,10.)(0.,){/Sine}{-1}
\FALabel(15.2447,14.2165)[br]{$\mathrm{W}$}
\FAProp(20.,10.)(15.5,6.5)(0.,){/Sine}{1}
\FALabel(17.2784,8.9935)[br]{$\mathrm{W}$}
\FAProp(20.,3.)(15.5,6.5)(0.,){/Sine}{0}
\FALabel(18.2216,5.4935)[bl]{$\gamma$}
\FAProp(5.5,10.)(11.5,10.)(0.,){/Sine}{0}
\FALabel(8.5,11.07)[b]{$\mathrm{Z}/\gamma$}
\FAProp(11.5,10.)(15.5,6.5)(0.,){/ScalarDash}{-1}
\FALabel(13.1239,7.75165)[tr]{$\phi^+$}
\FAVert(5.5,10.){0}
\FAVert(11.5,10.){0}
\FAVert(15.5,6.5){0}

\FADiagram{}
\FAProp(0.,15.)(5.5,10.)(0.,){/Straight}{1}
\FALabel(2.18736,11.8331)[tr]{$q$}
\FAProp(0.,5.)(5.5,10.)(0.,){/Straight}{-1}
\FALabel(3.31264,6.83309)[tl]{$\bar q$}
\FAProp(20.,17.)(11.5,10.)(0.,){/Sine}{-1}
\FALabel(15.2447,14.2165)[br]{$\mathrm{W}$}
\FAProp(20.,10.)(15.5,6.5)(0.,){/Sine}{1}
\FALabel(17.2784,8.9935)[br]{$\mathrm{W}$}
\FAProp(20.,3.)(15.5,6.5)(0.,){/Sine}{0}
\FALabel(18.2216,5.4935)[bl]{$\gamma$}
\FAProp(5.5,10.)(11.5,10.)(0.,){/Sine}{0}
\FALabel(8.5,11.07)[b]{$\mathrm{Z}/\gamma$}
\FAProp(11.5,10.)(15.5,6.5)(0.,){/Sine}{-1}
\FALabel(12.9593,7.56351)[tr]{$\mathrm{W}$}
\FAVert(5.5,10.){0}
\FAVert(11.5,10.){0}
\FAVert(15.5,6.5){0}

\FADiagram{}
\FAProp(0.,15.)(10.,14.5)(0.,){/Straight}{1}
\FALabel(5.0774,15.8181)[b]{$q$}
\FAProp(0.,5.)(10.,5.5)(0.,){/Straight}{-1}
\FALabel(5.0774,4.18193)[t]{$\bar q$}
\FAProp(20.,17.)(10.,5.5)(0.,){/Sine}{-1}
\FALabel(17.2913,15.184)[br]{$\mathrm{W}$}
\FAProp(20.,10.)(10.,14.5)(0.,){/Sine}{1}
\FALabel(17.8274,10.002)[tr]{$\mathrm{W}$}
\FAProp(20.,3.)(10.,10.)(0.,){/Sine}{0}
\FALabel(14.5911,5.71019)[tr]{$\gamma$}
\FAProp(10.,14.5)(10.,10.)(0.,){/Straight}{1}
\FALabel(9.18,12.25)[r]{$q'$}
\FAProp(10.,5.5)(10.,10.)(0.,){/Straight}{-1}
\FALabel(9.18,7.75)[r]{$q'$}
\FAVert(10.,14.5){0}
\FAVert(10.,5.5){0}
\FAVert(10.,10.){0}

\FADiagram{}
\FAProp(0.,15.)(10.,5.5)(0.,){/Straight}{1}
\FALabel(3.19219,13.2012)[bl]{$q$}
\FAProp(0.,5.)(10.,14.5)(0.,){/Straight}{-1}
\FALabel(3.10297,6.58835)[tl]{$\bar q$}
\FAProp(20.,17.)(10.,14.5)(0.,){/Sine}{-1}
\FALabel(14.6241,16.7737)[b]{$\mathrm{W}$}
\FAProp(20.,10.)(10.,10.)(0.,){/Sine}{1}
\FALabel(15.95,11.07)[b]{$\mathrm{W}$}
\FAProp(20.,3.)(10.,5.5)(0.,){/Sine}{0}
\FALabel(14.6241,3.22628)[t]{$\gamma$}
\FAProp(10.,5.5)(10.,10.)(0.,){/Straight}{-1}
\FALabel(10.82,7.75)[l]{$q$}
\FAProp(10.,14.5)(10.,10.)(0.,){/Straight}{1}
\FALabel(10.82,12.25)[l]{$q'$}
\FAVert(10.,5.5){0}
\FAVert(10.,14.5){0}
\FAVert(10.,10.){0}

\FADiagram{}
\FAProp(0.,15.)(10.,7.)(0.,){/Straight}{1}
\FALabel(2.82617,13.9152)[bl]{$q$}
\FAProp(0.,5.)(10.,14.5)(0.,){/Straight}{-1}
\FALabel(3.04219,6.54875)[tl]{$\bar q$}
\FAProp(20.,17.)(10.,14.5)(0.,){/Sine}{-1}
\FALabel(14.6241,16.7737)[b]{$\mathrm{W}$}
\FAProp(20.,10.)(15.5,6.5)(0.,){/Sine}{1}
\FALabel(17.2784,8.9935)[br]{$\mathrm{W}$}
\FAProp(20.,3.)(15.5,6.5)(0.,){/Sine}{0}
\FALabel(17.2784,4.0065)[tr]{$\gamma$}
\FAProp(10.,7.)(10.,14.5)(0.,){/Straight}{1}
\FALabel(10.82,10.75)[l]{$q'$}
\FAProp(10.,7.)(15.5,6.5)(0.,){/Sine}{0}
\FALabel(12.6097,5.68637)[t]{$\mathrm{W}$}
\FAVert(10.,7.){0}
\FAVert(10.,14.5){0}
\FAVert(15.5,6.5){0}

\FADiagram{}
\FAProp(0.,15.)(10.,14.5)(0.,){/Straight}{1}
\FALabel(5.0774,15.8181)[b]{$q$}
\FAProp(0.,5.)(10.,5.5)(0.,){/Straight}{-1}
\FALabel(5.0774,4.18193)[t]{$\bar q$}
\FAProp(20.,17.)(10.,10.)(0.,){/Sine}{-1}
\FALabel(17.4388,16.3076)[br]{$\mathrm{W}$}
\FAProp(20.,10.)(10.,14.5)(0.,){/Sine}{1}
\FALabel(17.2774,10.302)[tr]{$\mathrm{W}$}
\FAProp(20.,3.)(10.,5.5)(0.,){/Sine}{0}
\FALabel(14.6241,3.22628)[t]{$\gamma$}
\FAProp(10.,14.5)(10.,10.)(0.,){/Straight}{1}
\FALabel(9.18,12.25)[r]{$q'$}
\FAProp(10.,5.5)(10.,10.)(0.,){/Straight}{-1}
\FALabel(9.18,7.75)[r]{$q$}
\FAVert(10.,14.5){0}
\FAVert(10.,5.5){0}
\FAVert(10.,10.){0}

\FADiagram{}
\FAProp(0.,15.)(10.,14.5)(0.,){/Straight}{1}
\FALabel(5.0774,15.8181)[b]{$q$}
\FAProp(0.,5.)(10.,5.5)(0.,){/Straight}{-1}
\FALabel(5.0774,4.18193)[t]{$\bar q$}
\FAProp(20.,17.)(15.5,8.)(0.,){/Sine}{-1}
\FALabel(18.0564,15.0818)[br]{$\mathrm{W}$}
\FAProp(20.,10.)(10.,14.5)(0.,){/Sine}{1}
\FALabel(12.6226,14.248)[bl]{$\mathrm{W}$}
\FAProp(20.,3.)(15.5,8.)(0.,){/Sine}{0}
\FALabel(18.4221,6.0569)[bl]{$\gamma$}
\FAProp(10.,14.5)(10.,5.5)(0.,){/Straight}{1}
\FALabel(9.18,10.)[r]{$q'$}
\FAProp(10.,5.5)(15.5,8.)(0.,){/Sine}{0}
\FALabel(12.9114,5.81893)[tl]{$\mathrm{W}$}
\FAVert(10.,14.5){0}
\FAVert(10.,5.5){0}
\FAVert(15.5,8.){0}

\FADiagram{}
\FAProp(0.,15.)(10.,5.5)(0.,){/Straight}{1}
\FALabel(3.19219,13.2012)[bl]{$q$}
\FAProp(0.,5.)(10.,13.)(0.,){/Straight}{-1}
\FALabel(3.17617,6.28478)[tl]{$\bar q$}
\FAProp(20.,17.)(16.,13.5)(0.,){/Sine}{-1}
\FALabel(17.4593,15.9365)[br]{$\mathrm{W}$}
\FAProp(20.,10.)(16.,13.5)(0.,){/Sine}{1}
\FALabel(17.4593,11.0635)[tr]{$\mathrm{W}$}
\FAProp(20.,3.)(10.,5.5)(0.,){/Sine}{0}
\FALabel(14.6241,3.22628)[t]{$\gamma$}
\FAProp(10.,5.5)(10.,13.)(0.,){/Straight}{1}
\FALabel(10.82,9.25)[l]{$q$}
\FAProp(10.,13.)(16.,13.5)(0.,){/Sine}{0}
\FALabel(12.8713,14.3146)[b]{$\mathrm{Z}/\gamma$}
\FAVert(10.,5.5){0}
\FAVert(10.,13.){0}
\FAVert(16.,13.5){0}

\FADiagram{}
\FAProp(0.,15.)(10.,13.)(0.,){/Straight}{1}
\FALabel(5.30398,15.0399)[b]{$q$}
\FAProp(0.,5.)(10.,5.5)(0.,){/Straight}{-1}
\FALabel(5.0774,4.18193)[t]{$\bar q$}
\FAProp(20.,17.)(15.5,13.5)(0.,){/Sine}{-1}
\FALabel(17.2784,15.9935)[br]{$\mathrm{W}$}
\FAProp(20.,10.)(15.5,13.5)(0.,){/Sine}{1}
\FALabel(18.2216,12.4935)[bl]{$\mathrm{W}$}
\FAProp(20.,3.)(10.,5.5)(0.,){/Sine}{0}
\FALabel(15.3759,5.27372)[b]{$\gamma$}
\FAProp(10.,13.)(10.,5.5)(0.,){/Straight}{1}
\FALabel(9.18,9.25)[r]{$q$}
\FAProp(10.,13.)(15.5,13.5)(0.,){/Sine}{0}
\FALabel(12.8903,12.1864)[t]{$\mathrm{Z}/\gamma$}
\FAVert(10.,13.){0}
\FAVert(10.,5.5){0}
\FAVert(15.5,13.5){0}

\end{feynartspicture}
\end{small}
\end{center}

%% file: ppww_jhepstyle.bbl
\begin{thebibliography}{99}
%\cite{Chatrchyan:2011ns}
\bibitem{Chatrchyan:2011ns}
  S.~Chatrchyan {\it et al.}  [CMS Collaboration],
  %``Search for Resonances in the Dijet Mass Spectrum from 7 TeV pp Collisions at CMS,''
  Phys.\ Lett.\ B {\bf 704} (2011) 123
  [arXiv:1107.4771 [hep-ex]].
  %%CITATION = ARXIV:1107.4771;%%
%\cite{Aad:2011fc}
\bibitem{Aad:2011fc}
  G.~Aad {\it et al.}  [ATLAS Collaboration],
  %``Measurement of inclusive jet and dijet production in pp collisions at sqrt(s) = 7 TeV using the ATLAS detector,''
  arXiv:1112.6297 [hep-ex].
  %%CITATION = ARXIV:1112.6297;%%

%\cite{Ross:1975fq}
\bibitem{Ross:1975fq}
  D.~A.~Ross and M.~J.~G.~Veltman,
  %``Neutral Currents in Neutrino Experiments,''
  Nucl.\ Phys.\ B {\bf 95} (1975) 135.
  %%CITATION = NUPHA,B95,135;%%

%%%%%%%%%%%%%%% Sudakov Refs 8-11
%\cite{Kuhn:1999de}
\bibitem{Kuhn:1999de}
  J.~H.~K\"uhn and A.~A.~Penin,
  %``Sudakov logarithms in electroweak processes,''
  hep-ph/9906545.
  %%CITATION = HEP-PH/9906545;%%
%\cite{Fadin:1999bq}
\bibitem{Fadin:1999bq}
  V.~S.~Fadin, L.~N.~Lipatov, A.~D.~Martin and M.~Melles,
  %``Resummation of double logarithms in electroweak high-energy processes,''
  Phys.\ Rev.\ D {\bf 61} (2000) 094002
  [hep-ph/9910338].
  %%CITATION = HEP-PH/9910338;%%
%\cite{Ciafaloni:2000df}
\bibitem{Ciafaloni:2000df}
  M.~Ciafaloni, P.~Ciafaloni and D.~Comelli,
  %``Bloch-Nordsieck violating electroweak corrections to inclusive TeV scale hard processes,''
  Phys.\ Rev.\ Lett.\  {\bf 84} (2000) 4810
  [hep-ph/0001142].
  %%CITATION = HEP-PH/0001142;%%
%\cite{Kuhn:1999nn}
\bibitem{Kuhn:1999nn}
  J.~H.~K\"uhn, A.~A.~Penin and V.~A.~Smirnov,
  %``Summing up subleading Sudakov logarithms,''
  Eur.\ Phys.\ J.\ C {\bf 17} (2000) 97
  [hep-ph/9912503].
  %%CITATION = HEP-PH/9912503;%%
%\cite{Beenakker:2000kb}
\bibitem{Beenakker:2000kb}
  W.~Beenakker and A.~Werthenbach,
  %``New insights into the perturbative structure of electroweak Sudakov logarithms,''
  Phys.\ Lett.\ B {\bf 489} (2000) 148
  [hep-ph/0005316].
  %%CITATION = HEP-PH/0005316;%%
%\cite{Beenakker:2001kf}
\bibitem{Beenakker:2001kf}
  W.~Beenakker and A.~Werthenbach,
  %``Electroweak two loop Sudakov logarithms for on-shell fermions and bosons,''
  Nucl.\ Phys.\ B {\bf 630} (2002) 3
  [hep-ph/0112030].
  %%CITATION = HEP-PH/0112030;%%
%\cite{Melles:2000gw}
\bibitem{Melles:2000gw}
  M.~Melles,
  %``Subleading Sudakov logarithms in electroweak high-energy processes to all orders,''
  Phys.\ Rev.\ D {\bf 63} (2001) 034003
  [hep-ph/0004056].
  %%CITATION = HEP-PH/0004056;%%
%\cite{Melles:2000ia}
\bibitem{Melles:2000ia}
  M.~Melles,
  %``Resummation of Yukawa enhanced and subleading Sudakov logarithms in longitudinal gauge boson and Higgs production,''
  Phys.\ Rev.\ D {\bf 64} (2001) 014011
  [hep-ph/0012157].
  %%CITATION = HEP-PH/0012157;%%
%\cite{Denner:2003wi}
\bibitem{Denner:2003wi}
  A.~Denner, M.~Melles and S.~Pozzorini,
  %``Two loop electroweak angular dependent logarithms at high-energies,''
  Nucl.\ Phys.\ B {\bf 662} (2003) 299
  [hep-ph/0301241].
  %%CITATION = HEP-PH/0301241;%%
%\cite{Kuhn:2001hz}
\bibitem{Kuhn:2001hz}
  J.~H.~K\"uhn, S.~Moch, A.~A.~Penin and V.~A.~Smirnov,
  %``Next-to-next-to-leading logarithms in four fermion electroweak processes at high-energy,''
  Nucl.\ Phys.\ B {\bf 616} (2001) 286
   [Erratum-ibid.\ B {\bf 648} (2003) 455]
  [hep-ph/0106298].
  %%CITATION = HEP-PH/0106298;%%
%\cite{Feucht:2004rp}
\bibitem{Feucht:2004rp}
  B.~Feucht, J.~H.~K\"uhn, A.~A.~Penin, V.~A.~Smirnov,
  %``Two loop Sudakov form-factor in a theory with mass gap,''
  Phys.\ Rev.\ Lett.\  {\bf 93} (2004) 101802
  [hep-ph/0404082].
  %%CITATION = HEP-PH/0404082;%%
%\cite{Jantzen:2005xi}
\bibitem{Jantzen:2005xi}
  B.~Jantzen, J.~H.~K\"uhn, A.~A.~Penin, V.~A.~Smirnov,  and ,
  %``Two-loop electroweak logarithms,''
  Phys.\ Rev.\ D {\bf 72} (2005) 051301
   [Erratum-ibid.\ D {\bf 74} (2006) 019901]
  [hep-ph/0504111].
%\cite{Jantzen:2006jv}
\bibitem{Jantzen:2006jv}
  B.~Jantzen, V.~A.~Smirnov and ,
  %``The Two-loop vector form-factor in the Sudakov limit,''
  Eur.\ Phys.\ J.\ C {\bf 47} (2006) 671
  [hep-ph/0603133].
  %%CITATION = HEP-PH/0603133;%%
  %%CITATION = HEP-PH/0504111;%%
%\cite{Jantzen:2005az}
\bibitem{Jantzen:2005az}
  B.~Jantzen, J.~H.~K\"uhn, A.~A.~Penin, V.~A.~Smirnov and ,
  %``Two-loop electroweak logarithms in four-fermion processes at high energy,''
  Nucl.\ Phys.\ B {\bf 731} (2005) 188
   [Erratum-ibid.\ B {\bf 752} (2006) 327]
  [hep-ph/0509157].
  %%CITATION = HEP-PH/0509157;%%
%\cite{Denner:2000jv}
\bibitem{Denner:2000jv}
  A.~Denner and S.~Pozzorini,
  %``One loop leading logarithms in electroweak radiative corrections. 1. Results,''
  Eur.\ Phys.\ J.\ C {\bf 18} (2001) 461
  [hep-ph/0010201];
  %%CITATION = HEP-PH/0010201;%%
%\cite{Denner:2001gw}
%\bibitem{Denner:2001gw}
  A.~Denner and S.~Pozzorini,
  %``One loop leading logarithms in electroweak radiative corrections. 2. Factorization of collinear singularities,''
  Eur.\ Phys.\ J.\ C {\bf 21} (2001) 63
  [hep-ph/0104127];
  %%CITATION = HEP-PH/0104127;%%
%\cite{Pozzorini:2004rm}
%\bibitem{Pozzorini:2004rm}
  S.~Pozzorini,
  %``Next to leading mass singularities in two loop electroweak singlet form-factors,''
  Nucl.\ Phys.\ B {\bf 692} (2004) 135
  [hep-ph/0401087].
  %%CITATION = HEP-PH/0401087;%%
%\cite{Chiu:2009yx}
\bibitem{Chiu:2009yx}
  J.~-y.~Chiu, A.~Fuhrer, A.~H.~Hoang, R.~Kelley and A.~V.~Manohar,
  %``Soft-Collinear Factorization and Zero-Bin Subtractions,''
  Phys.\ Rev.\ D {\bf 79} (2009) 053007
  [arXiv:0901.1332 [hep-ph]].
  %%CITATION = ARXIV:0901.1332;%%

%\cite{Gribov:1966hs}
\bibitem{Gribov:1966hs}
  V.~N.~Gribov,
  %``Bremsstrahlung of hadrons at high energies,''
  Sov.\ J.\ Nucl.\ Phys.\  {\bf 5} (1967) 280
   [Yad.\ Fiz.\  {\bf 5} (1967) 399].
  %%CITATION = SJNCA,5,280;%%
%\cite{Gribov:1970ik}
\bibitem{Gribov:1970ik}
  V.~N.~Gribov, L.~N.~Lipatov and G.~V.~Frolov,
  %``The leading singularity in the j plane in quantum electrodynamics,''
  Sov.\ J.\ Nucl.\ Phys.\  {\bf 12} (1971) 543
   [Yad.\ Fiz.\  {\bf 12} (1970) 994].
  %%CITATION = SJNCA,12,543;%%
%\cite{Lipatov:1988ii}
\bibitem{Lipatov:1988ii}
  L.~N.~Lipatov,
  %``Massless Particle Bremsstrahlung Theorems For High-energy Hadron Interactions,''
  Nucl.\ Phys.\ B {\bf 307} (1988) 705.
  %%CITATION = NUPHA,B307,705;%%
%\cite{DelDuca:1990gz}
\bibitem{DelDuca:1990gz}
  V.~Del Duca,
  %``High-energy Bremsstrahlung Theorems For Soft Photons,''
  Nucl.\ Phys.\ B {\bf 345} (1990) 369.
  %%CITATION = NUPHA,B345,369;%%

%\cite{Kuhn:2007ca}
\bibitem{Kuhn:2007ca}
  J.~H.~K\"uhn, F.~Metzler and A.~A.~Penin,
  %``Next-to-next-to-leading electroweak logarithms in W-pair production at ILC,''
  Nucl.\ Phys.\ B {\bf 795} (2008) 277
   [Erratum-ibid.\  {\bf 818} (2009) 135]
  [arXiv:0709.4055 [hep-ph]].
  %%CITATION = ARXIV:0709.4055;%%
%\cite{Kuhn:2011mh}
\bibitem{Kuhn:2011mh}
  J.~H.~K\"uhn, F.~Metzler, A.~A.~Penin and S.~Uccirati,
  %``Next-to-Next-to-Leading Electroweak Logarithms for W-Pair Production at LHC,''
  JHEP {\bf 1106} (2011) 143
  [arXiv:1101.2563 [hep-ph]].
  %%CITATION = ARXIV:1101.2563;%%
%\cite{Fuhrer:2010eu}
\bibitem{Fuhrer:2010eu}
  A.~Fuhrer, A.~V.~Manohar, J.~-y.~Chiu and R.~Kelley,
  %``Radiative Corrections to Longitudinal and Transverse Gauge Boson and Higgs Production,''
  Phys.\ Rev.\ D {\bf 81} (2010) 093005
  [arXiv:1003.0025 [hep-ph]].
  %%CITATION = ARXIV:1003.0025;%%
%\cite{Kuhn:2006vh}
\bibitem{Kuhn:2006vh}
  J.~H.~K\"uhn, A.~Scharf and P.~Uwer,
  %``Electroweak effects in top-quark pair production at hadron colliders,''
  Eur.\ Phys.\ J.\ C {\bf 51} (2007) 37
  [hep-ph/0610335].
  %%CITATION = HEP-PH/0610335;%%
%\cite{Bernreuther:2006vg}
\bibitem{Bernreuther:2006vg}
  W.~Bernreuther, M.~Fuecker and Z.~-G.~Si,
  %``Weak interaction corrections to hadronic top quark pair production,''
  Phys.\ Rev.\ D {\bf 74} (2006) 113005
  [hep-ph/0610334].
  %%CITATION = HEP-PH/0610334;%%
%\cite{Moretti:2006nf}
\bibitem{Moretti:2006nf}
  S.~Moretti, M.~R.~Nolten and D.~A.~Ross,
  %``Weak corrections to gluon-induced top-antitop hadro-production,''
  Phys.\ Lett.\ B {\bf 639} (2006) 513
   [Erratum-ibid.\ B {\bf 660} (2008) 607]
  [hep-ph/0603083].
  %%CITATION = HEP-PH/0603083;%%
%\cite{Beenakker:1993yr}
\bibitem{Beenakker:1993yr}
  W.~Beenakker, A.~Denner, W.~Hollik, R.~Mertig, T.~Sack and D.~Wackeroth,
  %``Electroweak one loop contributions to top pair production in hadron colliders,''
  Nucl.\ Phys.\ B {\bf 411} (1994) 343.
  %%CITATION = NUPHA,B411,343;%%
%\cite{Kuhn:2009nf}
\bibitem{Kuhn:2009nf}
  J.~H.~K\"uhn, A.~Scharf and P.~Uwer,
  %``Weak effects in b-jet production at hadron colliders,''
  Phys.\ Rev.\ D {\bf 82} (2010) 013007
  [arXiv:0909.0059 [hep-ph]].
  %%CITATION = ARXIV:0909.0059;%%

%\cite{Ciccolini:2003jy}
\bibitem{Ciccolini:2003jy}
  M.~L.~Ciccolini, S.~Dittmaier and M.~Kramer,
  %``Electroweak radiative corrections to associated WH and ZH production at hadron colliders,''
  Phys.\ Rev.\ D {\bf 68} (2003) 073003
  [hep-ph/0306234].
  %%CITATION = HEP-PH/0306234;%%
%\cite{Denner:2011id}
\bibitem{Denner:2011id}
  A.~Denner, S.~Dittmaier, S.~Kallweit and A.~Muck,
  %``Electroweak corrections to Higgs-strahlung off W/Z bosons at the Tevatron and the LHC with HAWK,''
  JHEP {\bf 1203} (2012) 075
  [arXiv:1112.5142 [hep-ph]].
  %%CITATION = ARXIV:1112.5142;%%
%\cite{Ciccolini:2007ec}
\bibitem{Ciccolini:2007ec}
  M.~Ciccolini, A.~Denner and S.~Dittmaier,
  %``Electroweak and QCD corrections to Higgs production via vector-boson fusion at the LHC,''
  Phys.\ Rev.\ D {\bf 77} (2008) 013002
  [arXiv:0710.4749 [hep-ph]].
  %%CITATION = ARXIV:0710.4749;%%
%\cite{Actis:2008ug}
\bibitem{Actis:2008ug}
  S.~Actis, G.~Passarino, C.~Sturm and S.~Uccirati,
  %``NLO Electroweak Corrections to Higgs Boson Production at Hadron Colliders,''
  Phys.\ Lett.\ B {\bf 670} (2008) 12
  [arXiv:0809.1301 [hep-ph]].
  %%CITATION = ARXIV:0809.1301;%%

%\cite{Passarino:2007fp}
\bibitem{Passarino:2007fp}
  G.~Passarino, C.~Sturm and S.~Uccirati,
  %``Complete Two-Loop Corrections to H ---> gamma gamma,''
  Phys.\ Lett.\ B {\bf 655} (2007) 298
  [arXiv:0707.1401 [hep-ph]].
  %%CITATION = ARXIV:0707.1401;%%

 %\cite{Degrassi:2004mx}
% \bibitem{Degrassi:2004mx}
%   G.~Degrassi and F.~Maltoni,
%   %``Two-loop electroweak corrections to Higgs production at hadron colliders,''
%   Phys.\ Lett.\ B {\bf 600} (2004) 255
%   [hep-ph/0407249].
%   %%CITATION = HEP-PH/0407249;%%
% %\cite{Degrassi:2005mc}
% \bibitem{Degrassi:2005mc}
%   G.~Degrassi and F.~Maltoni,
%   %``Two-loop electroweak corrections to the Higgs-boson decay H ---> gamma gamma,''
%   Nucl.\ Phys.\ B {\bf 724} (2005) 183
%   [hep-ph/0504137].
%   %%CITATION = HEP-PH/0504137;%%
%\cite{Kniehl:1990mq}
\bibitem{Kniehl:1990mq}
  B.~A.~Kniehl,
  %``Radiative corrections for $H \to Z Z$ in the standard model,''
  Nucl.\ Phys.\ B {\bf 352} (1991) 1.
  %%CITATION = NUPHA,B352,1;%%
%\cite{Kniehl:1991xe}
\bibitem{Kniehl:1991xe}
  B.~A.~Kniehl,
  %``Radiative corrections for $H \to W^{+} W^{-}$ ($\gamma$) in the standard model,''
  Nucl.\ Phys.\ B {\bf 357} (1991) 439.
  %%CITATION = NUPHA,B357,439;%%
%\cite{Kniehl:1991ze}
\bibitem{Kniehl:1991ze}
  B.~A.~Kniehl,
  %``Radiative corrections for $H \to$ f anti-f ($\gamma$) in the standard model,''
  Nucl.\ Phys.\ B {\bf 376} (1992) 3.
  %%CITATION = NUPHA,B376,3;%%
%\cite{Moretti:2006ea}
\bibitem{Moretti:2006ea}
  S.~Moretti, M.~R.~Nolten and D.~A.~Ross,
  %``Weak corrections to four-parton processes,''
  Nucl.\ Phys.\ B {\bf 759} (2006) 50
  [hep-ph/0606201].
  %%CITATION = HEP-PH/0606201;%%
%\cite{Dittmaier:2012kx}
\bibitem{Dittmaier:2012kx}
  S.~Dittmaier, A.~Huss and C.~Speckner,
  %``Weak radiative corrections to dijet production at hadron colliders,''
  arXiv:1210.0438 [hep-ph].
  %%CITATION = ARXIV:1210.0438;%%
%\cite{Kasprzik:2011ds}
\bibitem{Kasprzik:2011ds}
  T.~Kasprzik,
  %``Vector-boson production at the LHC: QCD and electroweak effects,''
  PoS EPS {\bf -HEP2011} (2011) 358
  [arXiv:1111.1146 [hep-ph]].
  %%CITATION = ARXIV:1111.1146;%%
%\cite{Kuhn:2004em}
\bibitem{Kuhn:2004em}
  J.~H.~K\"uhn, A.~Kulesza, S.~Pozzorini and M.~Schulze,
  %``Logarithmic electroweak corrections to hadronic Z+1 jet production at large transverse momentum,''
  Phys.\ Lett.\ B {\bf 609} (2005) 277
  [hep-ph/0408308].
  %%CITATION = HEP-PH/0408308;%%

%\cite{Kuhn:2005az}
\bibitem{Kuhn:2005az}
  J.~H.~K\"uhn, A.~Kulesza, S.~Pozzorini and M.~Schulze,
  %``One-loop weak corrections to hadronic production of Z bosons at large transverse momenta,''
  Nucl.\ Phys.\ B {\bf 727} (2005) 368
  [hep-ph/0507178].
  %%CITATION = HEP-PH/0507178;%%
%\cite{Kuhn:2005gv}
\bibitem{Kuhn:2005gv}
  J.~H.~K\"uhn, A.~Kulesza, S.~Pozzorini and M.~Schulze,
  %``Electroweak corrections to hadronic photon production at large transverse momenta,''
  JHEP {\bf 0603} (2006) 059
  [hep-ph/0508253].
  %%CITATION = HEP-PH/0508253;%%
%\cite{Kuhn:2007qc}
\bibitem{Kuhn:2007qc}
  J.~H.~K\"uhn, A.~Kulesza, S.~Pozzorini and M.~Schulze,
  %``Electroweak corrections to large transverse momentum production of W bosons at the LHC,''
  Phys.\ Lett.\ B {\bf 651} (2007) 160
  [hep-ph/0703283 [HEP-PH]].
  %%CITATION = HEP-PH/0703283;%%
%\cite{Kuhn:2007cv}
\bibitem{Kuhn:2007cv}
  J.~H.~K\"uhn, A.~Kulesza, S.~Pozzorini and M.~Schulze,
  %``Electroweak corrections to hadronic production of W bosons at large transverse momenta,''
  Nucl.\ Phys.\ B {\bf 797} (2008) 27
  [arXiv:0708.0476 [hep-ph]].
  %%CITATION = ARXIV:0708.0476;%%
%\cite{Hollik:2007sq}
\bibitem{Hollik:2007sq}
  W.~Hollik, T.~Kasprzik and B.~A.~Kniehl,
  %``Electroweak corrections to W-boson hadroproduction at finite transverse momentum,''
  Nucl.\ Phys.\ B {\bf 790} (2008) 138
  [arXiv:0707.2553 [hep-ph]].
  %%CITATION = ARXIV:0707.2553;%%
%\cite{Denner:2009gj}
\bibitem{Denner:2009gj}
  A.~Denner, S.~Dittmaier, T.~Kasprzik and A.~M\"uck,
  %``Electroweak corrections to W + jet hadroproduction including leptonic W-boson decays,''
  JHEP {\bf 0908} (2009) 075
  [arXiv:0906.1656 [hep-ph]].
  %%CITATION = ARXIV:0906.1656;%%
%\cite{Denner:2011vu}
\bibitem{Denner:2011vu}
  A.~Denner, S.~Dittmaier, T.~Kasprzik and A.~M\"uck,
  %``Electroweak corrections to dilepton + jet production at hadron colliders,''
  JHEP {\bf 1106} (2011) 069
  [arXiv:1103.0914 [hep-ph]].
  %%CITATION = ARXIV:1103.0914;%%
%\cite{Accomando:2001fn}
\bibitem{Accomando:2001fn}
  E.~Accomando, A.~Denner and S.~Pozzorini,
  %``Electroweak correction effects in gauge boson pair production at the CERN LHC,''
  Phys.\ Rev.\ D {\bf 65} (2002) 073003
  [hep-ph/0110114].
  %%CITATION = HEP-PH/0110114;%%
%\cite{Accomando:2004de}
\bibitem{Accomando:2004de}
  E.~Accomando, A.~Denner and A.~Kaiser,
  %``Logarithmic electroweak corrections to gauge-boson pair production at the LHC,''
  Nucl.\ Phys.\ B {\bf 706} (2005) 325
  [hep-ph/0409247].
%\cite{Accomando:2005ra}
\bibitem{Accomando:2005ra}
  E.~Accomando, A.~Denner, C.~Meier,
  %``Electroweak corrections to $W \gamma$ and $Z \gamma$ production at the LHC,''
  Eur.\ Phys.\ J.\  {\bf C47 } (2006)  125-146.
  [hep-ph/0509234].
%\cite{Ohnemus:1991gb}
\bibitem{Ohnemus:1991gb}
  J.~Ohnemus,
  %``An Order $\alpha^- s$ calculation of hadronic $W^\pm Z$ production,''
  Phys.\ Rev.\  {\bf D44 } (1991)  3477-3489.
%\cite{Frixione:1993yp}
\bibitem{Frixione:1993yp}
  S.~Frixione,
  %``A Next-to-leading order calculation of the cross-section for the production of W+ W- pairs in hadronic collisions,''
  Nucl.\ Phys.\  {\bf B410 } (1993)  280-324.
%\cite{Ohnemus:1994qp}
\bibitem{Ohnemus:1994qp}
  J.~Ohnemus,
  %``Hadronic $Z \gamma$ production with QCD corrections and leptonic decays,''
  Phys.\ Rev.\  {\bf D51 } (1995)  1068-1076.
  [hep-ph/9407370].
%\cite{Ohnemus:1994ff}
\bibitem{Ohnemus:1994ff} J.~Ohnemus,
  % ``Hadronic $Z Z$, $W^{-} W^{+}$, and $W^\pm Z$ production with QCD
  % corrections and leptonic decays,''
  Phys.\ Rev.\ {\bf D50 } (1994) 1931-1945.  [arXiv:hep-ph/9403331 [hep-ph]].
%\cite{Dixon:1998py}
\bibitem{Dixon:1998py}
  L.~J.~Dixon, Z.~Kunszt, A.~Signer,
  % ``Helicity amplitudes for O(alpha-s) production of $W^{+} W^{-}$,
  % $W^\pm Z$, $Z Z$, $W^\pm \gamma$, or $Z \gamma$ pairs at hadron
  % colliders,''
  Nucl.\ Phys.\  {\bf B531 } (1998)  3-23.
  [hep-ph/9803250].  
%\cite{Dixon:1999di}
\bibitem{Dixon:1999di}
  L.~J.~Dixon, Z.~Kunszt, A.~Signer,
  % ``Vector boson pair production in hadronic collisions at order
  % $\alpha_s$ : Lepton correlations and anomalous couplings,''
  Phys.\ Rev.\  {\bf D60 } (1999)  114037.
  [hep-ph/9907305].
%\cite{DeFlorian:2000sg}
\bibitem{DeFlorian:2000sg}
  D.~De Florian, A.~Signer,
  %``W gamma and Z gamma production at hadron colliders,''
  Eur.\ Phys.\ J.\  {\bf C16 } (2000)  105-114.
  [hep-ph/0002138].
%\cite{Campbell:1999ah}
\bibitem{Campbell:1999ah}
  J.~M.~Campbell, R.~K.~Ellis,
  %``An Update on vector boson pair production at hadron colliders,''
  Phys.\ Rev.\  {\bf D60 } (1999)  113006.
  [hep-ph/9905386].
%\cite{Campbell:2011bn}
\bibitem{Campbell:2011bn}
  J.~M.~Campbell, R.~K.~Ellis and C.~Williams,
  %``Vector boson pair production at the LHC,''
  JHEP {\bf 1107} (2011) 018
  [arXiv:1105.0020 [hep-ph]].
  %%CITATION = ARXIV:1105.0020;%%
%\cite{Frixione:2006he}
\bibitem{Frixione:2006he}
  S.~Frixione, B.~R.~Webber,
  %``The MC@NLO 3.2 event generator,''
  [hep-ph/0601192].
%\cite{Glover:1988rg}
\bibitem{Glover:1988rg}
  E.~W.~N.~Glover, J.~J.~van der Bij,
  %``Z Boson Pair Production Via Gluon Fusion,''
  Nucl.\ Phys.\  {\bf B321 } (1989)  561.
%\cite{Kao:1990tt}
\bibitem{Kao:1990tt}
  C.~Kao, D.~A.~Dicus,
  %``Production of W+ W- from gluon fusion,''
  Phys.\ Rev.\  {\bf D43 } (1991)  1555-1559.
%\cite{Duhrssen:2005bz}
\bibitem{Duhrssen:2005bz}
  M.~Duhrssen, K.~Jakobs, J.~J.~van der Bij, P.~Marquard,
  %``The Process gg ---> WW as a background to the Higgs signal at the LHC,''
  JHEP {\bf 0505 } (2005)  064.
  [hep-ph/0504006].
%\cite{Binoth:2006mf}
\bibitem{Binoth:2006mf}
  T.~Binoth, M.~Ciccolini, N.~Kauer, M.~Kr\"amer,
  %``Gluon-induced W-boson pair production at the LHC,''
  JHEP {\bf 0612 } (2006)  046.
  [hep-ph/0611170].
%\cite{Campbell:2011cu}
\bibitem{Campbell:2011cu}
  J.~M.~Campbell, R.~K.~Ellis and C.~Williams,
  %``Gluon-Gluon Contributions to W+ W- Production and Higgs Interference Effects,''
  JHEP {\bf 1110} (2011) 005
  [arXiv:1107.5569 [hep-ph]].
  %%CITATION = ARXIV:1107.5569;%%

%\cite{Lemoine:1979pm}
\bibitem{Lemoine:1979pm}
  M.~Lemoine and M.~J.~G.~Veltman,
  %``Radiative Corrections to e+ e- ---> W+ W- in the Weinberg Model,''
  Nucl.\ Phys.\ B {\bf 164} (1980) 445.
  %%CITATION = NUPHA,B164,445;%%
%\cite{Bohm:1987ck}
\bibitem{Bohm:1987ck}
  M.~B\"ohm, A.~Denner, T.~Sack, W.~Beenakker, F.~A.~Berends and H.~Kuijf,
  %``Electroweak Radiative Corrections to e+ e- ---> W+ W-,''
  Nucl.\ Phys.\ B {\bf 304} (1988) 463.
  %%CITATION = NUPHA,B304,463;%%
%\cite{Fleischer:1987xa}
\bibitem{Fleischer:1987xa}
  J.~Fleischer, F.~Jegerlehner and M.~Zralek,
  %``Radiative Corrections To Helicity Amplitudes For The Process E+ E- ---> W+ W-,''
  BI-TP-88/03.
  %%CITATION = BI-TP-88/03;%%
%\cite{Beenakker:1993tt}
\bibitem{Beenakker:1993tt}
  W.~Beenakker, A.~Denner, S.~Dittmaier, R.~Mertig and T.~Sack,
  %``High-energy approximation for on-shell W pair production,''
  Nucl.\ Phys.\ B {\bf 410} (1993) 245.
  %%CITATION = NUPHA,B410,245;%%
%\cite{Beenakker:1998gr}
\bibitem{Beenakker:1998gr}
  W.~Beenakker, F.~A.~Berends and A.~P.~Chapovsky,
  %``Radiative corrections to pair production of unstable particles: results for e+ e- ---> four fermions,''
  Nucl.\ Phys.\ B {\bf 548} (1999) 3
  [hep-ph/9811481].
  %%CITATION = HEP-PH/9811481;%%
%\cite{Jadach:2001mp}
\bibitem{Jadach:2001mp}
  S.~Jadach, W.~Placzek, M.~Skrzypek, B.~F.~L.~Ward and Z.~Was,
  %``The Monte Carlo program KoralW version 1.51 and the concurrent Monte Carlo KoralW and YFSWW3 with all background graphs and first order corrections to W pair production,''
  Comput.\ Phys.\ Commun.\  {\bf 140} (2001) 475
  [hep-ph/0104049].
  %%CITATION = HEP-PH/0104049;%%
%\cite{Denner:2000bj}
\bibitem{Denner:2000bj}
  A.~Denner, S.~Dittmaier, M.~Roth and D.~Wackeroth,
  %``Electroweak radiative corrections to e+ e- ---> W W ---> 4 fermions in double pole approximation: The RACOONWW approach,''
  Nucl.\ Phys.\ B {\bf 587} (2000) 67
  [hep-ph/0006307];
  %%CITATION = HEP-PH/0006307;%%
%\cite{Denner:2002cg}
%\bibitem{Denner:2002cg}
  A.~Denner, S.~Dittmaier, M.~Roth and D.~Wackeroth,
  %``RACOONWW1.3: A Monte Carlo program for four fermion production at e+ e- colliders,''
  Comput.\ Phys.\ Commun.\  {\bf 153} (2003) 462
  [hep-ph/0209330].
  %%CITATION = HEP-PH/0209330;%%
%\cite{Denner:2005es}
\bibitem{Denner:2005es}
  A.~Denner, S.~Dittmaier, M.~Roth and L.~H.~Wieders,
  %``Complete electroweak O(alpha) corrections to charged-current e+e- ---> 4 fermion processes,''
  Phys.\ Lett.\ B {\bf 612} (2005) 223
   [Erratum-ibid.\ B {\bf 704} (2011) 667]
  [hep-ph/0502063];
  %%CITATION = HEP-PH/0502063;%%
%\cite{Denner:2005fg}
%\bibitem{Denner:2005fg}
  A.~Denner, S.~Dittmaier, M.~Roth and L.~H.~Wieders,
  %``Electroweak corrections to charged-current e+ e- ---> 4 fermion processes: Technical details and further results,''
  Nucl.\ Phys.\ B {\bf 724} (2005) 247
   [Erratum-ibid.\ B {\bf 854} (2012) 504]
  [hep-ph/0505042].
  %%CITATION = HEP-PH/0505042;%%
%\cite{Martin:2004dh}
\bibitem{Martin:2004dh}
  A.~D.~Martin, R.~G.~Roberts, W.~J.~Stirling, R.~S.~Thorne,
  %``Parton distributions incorporating QED contributions,''
  Eur.\ Phys.\ J.\  {\bf C39 } (2005)  155-161.
  [hep-ph/0411040].
%\cite{Denner:1995jv}
\bibitem{Denner:1995jv}
  A.~Denner, S.~Dittmaier and R.~Schuster,
  %``Radiative corrections to $\gamma \gamma \to W^{+} W^{-}$ in the electroweak standard model,''
  Nucl.\ Phys.\ B {\bf 452} (1995) 80
  [hep-ph/9503442].
  %%CITATION = HEP-PH/9503442;%%
%\cite{Archibald:2008zzb}
\bibitem{Archibald:2008zzb}
  J.~Archibald, S.~Hoeche, F.~Krauss, F.~Siegert, T.~Gleisberg, M.~Schonherr, S.~Schumann and J.~-C.~Winter,
  %``Simulation of photon-photon interactions in hadron collisions with SHERPA,''
  Nucl.\ Phys.\ Proc.\ Suppl.\  {\bf 179-180} (2008) 218.
  %%CITATION = NUPHZ,179-180,218;%%
%\cite{Pierzchala:2008xc}
\bibitem{Pierzchala:2008xc}
  T.~Pierzchala and K.~Piotrzkowski,
  %``Sensitivity to anomalous quartic gauge couplings in photon-photon interactions at the LHC,''
  Nucl.\ Phys.\ Proc.\ Suppl.\  {\bf 179-180} (2008) 257
  [arXiv:0807.1121 [hep-ph]].
  %%CITATION = ARXIV:0807.1121;%%

%\cite{Accomando:2005xp}
\bibitem{Accomando:2005xp}
  E.~Accomando and A.~Kaiser,
  %``Electroweak corrections and anomalous triple gauge-boson couplings in $W^{+} W^{-}$ and $W^\pm Z$ production at the LHC,''
  Phys.\ Rev.\ D {\bf 73} (2006) 093006
  [hep-ph/0511088].
  %%CITATION = HEP-PH/0511088;%%
%\cite{Roth:2004ti}
\bibitem{Roth:2004ti}
  M.~Roth and S.~Weinzierl,
  %``QED corrections to the evolution of parton distributions,''
  Phys.\ Lett.\ B {\bf 590} (2004) 190
  [hep-ph/0403200].
  %%CITATION = HEP-PH/0403200;%%
\bibitem{Kublbeck:1990xc}
  J.~K\"ublbeck, M.~B\"ohm and A.~Denner,
  %``Feyn Arts: Computer Algebraic Generation Of Feynman Graphs And Amplitudes,''
  Comput.\ Phys.\ Commun.\  {\bf 60} (1990) 165;\\
  %%CITATION = CPHCB,60,165;%%
  H.~Eck and J.~K\"ublbeck, {\it Guide to FeynArts 1.0\/},
  University of W\"urzburg, 1992.
\bibitem{Hahn:2000kx}
  T.~Hahn,
  %``Generating Feynman diagrams and amplitudes with FeynArts 3,''
  Comput.\ Phys.\ Commun.\  {\bf 140} (2001) 418
  [hep-ph/0012260].
\bibitem{Hahn:1998yk}
  T.~Hahn and M.~P\'erez-Victoria,
  %``Automatized one-loop calculations in four and D dimensions,''
  Comput.\ Phys.\ Commun.\  {\bf 118} (1999) 153
  [hep-ph/9807565].

%\cite{Hahn:2001rv}
\bibitem{Hahn:2001rv}
  T.~Hahn, C.~Schappacher,
  %``The Implementation of the minimal supersymmetric standard model in FeynArts and FormCalc,''
  Comput.\ Phys.\ Commun.\  {\bf 143 } (2002)  54-68.
  [hep-ph/0105349].
%\cite{Nogueira:1991ex}
\bibitem{Nogueira:1991ex}
  Nogueira, P.
  \newblock {Automatic Feynman graph generation}.
  \newblock J. Comput. Phys., \textbf{vol. 105 (1993)} pp. 279--289.
%\cite{Vermaseren:2000nd}
\bibitem{Vermaseren:2000nd}
Vermaseren, J.A.M.
\newblock {New features of FORM}.
\newblock  (2000).
%\newblock \eprint{math-ph/0010025}. 
  %%CITATION = HEP-PH 9807565;%%
%\cite{vanOldenborgh:1989wn}
%\cite{Passarino:1978jh}
\bibitem{Passarino:1978jh}
  G.~Passarino, M.~J.~G.~Veltman,
  %``One Loop Corrections for e+ e- Annihilation Into mu+ mu- in the Weinberg Model,''
  Nucl.\ Phys.\  {\bf B160 } (1979)  151.
\bibitem{vanOldenborgh:1989wn}
  G.~J.~van Oldenborgh, J.~A.~M.~Vermaseren,
  %``New Algorithms for One Loop Integrals,''
  Z.\ Phys.\  {\bf C46 } (1990)  425-438.
%\cite{Denner:1991kt}
\bibitem{Denner:1991kt}
  A.~Denner,
  %``Techniques for calculation of electroweak radiative corrections at the one loop level and results for W physics at LEP-200,''
  Fortsch.\ Phys.\  {\bf 41 } (1993)  307-420.
  [arXiv:0709.1075 [hep-ph]].
%\cite{Dittmaier:2001ay}
\bibitem{Dittmaier:2001ay}
  S.~Dittmaier, M.~Kr\"amer, 1,
  %``Electroweak radiative corrections to W boson production at hadron colliders,''
  Phys.\ Rev.\  {\bf D65 } (2002)  073007.
  [hep-ph/0109062].
%%\cite{Dittmaier:1998nn}
%\bibitem{Dittmaier:1998nn}
%  S.~Dittmaier,
%  %``Weyl-van-der-Waerden formalism for helicity amplitudes of massive
%  %particles,''
%  Phys.\ Rev.\  D {\bf 59} (1999) 016007
%  [hep-ph/9805445].
%  %%CITATION = PHRVA,D59,016007;%%
%\cite{Mertig:1990an}
\bibitem{Mertig:1990an}
  R.~Mertig, M.~B\"ohm and A.~Denner,
  %``FEYN CALC: Computer algebraic calculation of Feynman amplitudes,''
  Comput.\ Phys.\ Commun.\  {\bf 64} (1991) 345.
  %%CITATION = CPHCB,64,345;%%
%\cite{Alwall:2007st}
\bibitem{Alwall:2007st}
  J.~Alwall {\it et al.},
  %``MadGraph/MadEvent v4: The New Web Generation,''
  JHEP {\bf 0709 } (2007)  028.
  [arXiv:0706.2334 [hep-ph]].

%\cite{Lepage:1977sw}
\bibitem{Lepage:1977sw}
G.~P.~Lepage,
%``A New Algorithm For Adaptive Multidimensional Integration,''
J.\ Comput.\ Phys.\  {\bf 27} (1978) 192 and
%%CITATION = JCTPA,27,192;%%
%\cite{Lepage:1980dq}
%\bibitem{Lepage:1980dq}
%G.~P.~Lepage,
%``Vegas: An Adaptive Multidimensional Integration Program,''
CLNS-80/447.
%\cite{Harris:2001sx}
\bibitem{Harris:2001sx}
  B.~W.~Harris and J.~F.~Owens,
  %``The Two cutoff phase space slicing method,''
  Phys.\ Rev.\ D {\bf 65} (2002) 094032
  [hep-ph/0102128].
  %%CITATION = HEP-PH/0102128;%%
\bibitem{Dittmaier:1999mb}
  S.~Dittmaier,
  %``A general approach to photon radiation off fermions,''
  Nucl.\ Phys.\ B {\bf 565} (2000) 69
  [hep-ph/9904440].
  %%CITATION = HEP-PH 9904440;%%
%\cite{Bloch:1937pw}
\bibitem{Bloch:1937pw}
  F.~Bloch, A.~Nordsieck,
  %``Note on the Radiation Field of the electron,''
  Phys.\ Rev.\  {\bf 52 } (1937)  54-59.
%\cite{Diener:2003ss}
\bibitem{Diener:2003ss}
  K.~P.~O.~Diener, S.~Dittmaier and W.~Hollik,
  %``Electroweak radiative corrections to deep inelastic neutrino scattering: Implications for NuTeV?,''
  Phys.\ Rev.\ D {\bf 69} (2004) 073005
  [hep-ph/0310364].
  %%CITATION = HEP-PH/0310364;%%
%\href{http://www.slac.stanford.edu/spires/find/hep/www?r=clns-80\%2F447}{SPIRES
%entry} 
%\cite{Rubin:2010xp}
\bibitem{Rubin:2010xp}
  M.~Rubin, G.~P.~Salam and S.~Sapeta,
  %``Giant QCD K-factors beyond NLO,''
  JHEP {\bf 1009} (2010) 084
  [arXiv:1006.2144 [hep-ph]].
  %%CITATION = ARXIV:1006.2144;%%
%\bibitem{Amsler:2008zzb}
%  C.~Amsler {\it et al.}  [Particle Data Group],
%  %``Review of particle physics,''
%  Phys.\ Lett.\  B {\bf 667} (2008) 1.
%  %%CITATION = PHLTA,B667,1;%% 
%\cite{Martin:2009iq}
\bibitem{Martin:2009iq}
  A.~D.~Martin {\it et al.},
  %``Parton distributions for the LHC,''
  Eur.\ Phys.\ J.\  {\bf C63}, (2009) 189-285.
  [arXiv:0901.0002 [hep-ph]].
%\cite{Whalley:2005nh}
\bibitem{Whalley:2005nh}
  M.~R.~Whalley, D.~Bourilkov and R.~C.~Group, in {\sl HERA and the
    LHC}, eds.\ A. de Roeck and H. Jung (CERN-2005-014, Geneva, 2005), p.~575,
  %``The Les Houches Accord PDFs (LHAPDF) and Lhaglue,''
  hep-ph/0508110.
  %%CITATION = HEP-PH/0508110;%%
%\cite{Dittmaier:2009cr}
\bibitem{Dittmaier:2009cr}
  S.~Dittmaier and M.~Huber,
  %``Radiative corrections to the neutral-current Drell-Yan process in the Standard Model and its minimal supersymmetric extension,''
  JHEP {\bf 1001} (2010) 060
  [arXiv:0911.2329 [hep-ph]].
  %%CITATION = ARXIV:0911.2329;%%


\end{thebibliography}
